\documentclass[aps,prb,10pt,twocolumn,notitlepage,superscriptaddress,nobalancelastpage,raggedbottom,longbibliography]{revtex4-2}

\usepackage{amsmath,amssymb,graphicx,leftindex}
\usepackage[caption=false]{subfig}
\usepackage[version=4]{mhchem}
\usepackage{tensor}

\usepackage{xcolor}
\usepackage{bm}
\usepackage{comment}
\usepackage[normalem]{ulem}
\usepackage{hyperref}
\definecolor{jacksonsPurple}{RGB}{45,48,145}
\hypersetup{
    breaklinks=true,
    colorlinks=true,
	linkcolor={jacksonsPurple},
	citecolor={jacksonsPurple},
	urlcolor={jacksonsPurple}
}

\newcommand{\wIn}{\omega_{\textrm{in}}}

\newcommand{\wIC}{\omega_{\textrm{ic}}}

\newcommand{\Dop}{\hat{\bm{D}}}
\newcommand{\GTen}{\hat{\bm{\mathcal{G}}}}
\newcommand{\Gop}{\hat{\mathcal{G}}}

\newcommand{\polVec}{\vec{\mathbf{e}}}
\newcommand{\polVecC}{\vec{\mathbf{e}}^\dag}
\newcommand{\polTensor}{\vec{\boldsymbol{\epsilon}}}
\newcommand{\polTensorC}{\vec{\boldsymbol{\epsilon}}^\dag}

\newcommand{\seeAppendixX}[1]{[see Appendix #1]}

\newcommand{\Ham}{\hat{H}}
\newcommand{\HamCav}{\hat{H}_{\rm{cav}}}
\newcommand{\HamCavOne}{\hat{\bm{H}}_{\rm{cav,1}}}
\newcommand{\HamCavTwo}{\hat{\bm{H}}_{\rm{cav,2}}}
\newcommand{\HamMat}{\hat{H}_{\rm{mat}}}
\newcommand{\DMat}[1]{\hat{D}^{#1}}

\newcommand{\bra}[1]{\left<#1\right|}
\newcommand{\ket}[1]{\left|#1\right>}
\newcommand{\OPc}[2]{\hat{#1}_{#2}^{\dag}}
\newcommand{\OP}[2]{\hat{#1}_{#2}^{\vphantom{\dag}}}

\newcommand{\A}[1]{\OP{a}{#1}}
\newcommand{\AD}[1]{\OPc{a}{#1}}

\newcommand{\BinX}[1]{\OP{b}{\textrm{in},#1}}
\newcommand{\BoutXD}[1]{\OPc{b}{\textrm{out},#1}}
\newcommand{\BoutX}[1]{\OP{b}{\textrm{out},#1}}

\begin{document}

\title{Polarization-Resolved Photon Statistics of Cavity Quantum Materials}

\author{Benjamin Kass}
\affiliation{Department of Physics and Astronomy, University of Pennsylvania, Philadelphia, PA 19104, USA}
\author{Spenser Talkington}
\affiliation{Department of Physics and Astronomy, University of Pennsylvania, Philadelphia, PA 19104, USA}
\author{Martin Claassen}
\affiliation{Department of Physics and Astronomy, University of Pennsylvania, Philadelphia, PA 19104, USA}
\date{\today}

\begin{abstract}
By forming hybrid light-matter states, optical cavities offer a route for engineering material properties, however, unambiguously probing the effects of light-matter coupling remains difficult. Here, we show that the polarization-resolved statistics of photons transmitted through a cavity, measurable via $g^{(2)}$, provide one such diagnostic. By relating $g^{(2)}$ to matter correlation functions such as the Raman structure factor, we link photon bunching and antibunching to material properties. By applying this method to the stripy-to-antiferromagnetic transition in the Kitaev-Heisenberg spin model, we find that polarization-dependent patterns of bunching and antibunching encode the magnetic point-group symmetries of each phase and characterize the behavior at the phase boundary. Finally, we predict measuring $g^{(2)}$ for output photon pairs polarized orthogonal to the input field will isolate higher-order light-matter scattering processes that probe higher-order material correlations.
\end{abstract}

\maketitle

\section{Introduction}

    Cavity quantum-electrodynamical engineering of matter phases has emerged as a promising area of research, with proposed routes to controlling a wide range of quantum materials \cite{bloch22,schlawin21,hubener21,hubener24,garciavidal21,bretscher2026fluctuation}. By forming hybrid light-matter states, optical cavities can modify material properties and potentially drive quantum materials into new phases, including quantum Hall states \cite{appugliese2022breakdown,enkner2025tunable,graziotto2025cavity}, modified superconductivity \cite{sentef2018cavity,schlawin2019cavity,grankin2021enhancement,curtis2019cavity,thomas2025exploring,keren2026cavity,xu2026vacuum}, metal-insulator transitions \cite{jarc2023cavity,fassioli2025controlling}, quantum criticality \cite{weber2023cavity,kass2024,sur2025amplified}, spin liquids \cite{Chiocchetta2021,vinas2023controlling,nambiar2025diagnosing} and more \cite{orgiu15,ebbesen16,juraschek19,mazza19,kiffner2019manipulating,nagarajan21,latini21,eckhardt2022quantum,passetti23,shaffer2024entanglement,kipp2024}. However, determining whether and by which mechanism photons are responsible for changes to matter properties remains a key challenge. Toward this end, it is important to develop techniques for detecting quantum-electrodynamically induced changes in the state of matter in cavity-embedded materials.

    Higher-order photon correlations, such as the second-order photon correlation function $g^{(2)}(t)$, a standard quantum-optical tool for characterizing quantum emitters \cite{mandel1959fluctuations,glauber1963quantum,brown1956correlation,paul1982photon,chang2014quantum}, provide one natural route toward such a diagnostic for cavity materials. Conventional transmission and Raman spectra are often insufficient to diagnose cavity-induced modifications of a matter phase, as they can only reveal polaritonic splittings and selection rules of single excitations without directly probing photon-fluctuation-induced changes to the material \cite{ridolfo2012photon,flayac2013input,goto2019figure,heinisch2026high,trivedi2019photon,chen2022photon,delteil2019towards,kass2024,talkington2026}. To measure the effect of photons on the material, at minimum a second photon is required, necessitating higher-order correlation functions like $g^{(2)}(t)$.

    Recent works have investigated higher-order photon correlation functions and quadrature measurements as a dictionary to probe properties of quantum materials with and without cavities \cite{kass2024,nambiar2025diagnosing,grunwald2025cavity}. In a recent paper \cite{kass2024}, we showed that the photon statistics of light transmitted through an optical cavity exhibit pronounced antibunching for cavity quantum materials near a zero-temperature quantum critical point (QCP). This effect originates from cavity-photon induced changes to the matter phase which becomes particularly significant near a QCP; or, equivalently, from an effective photonic nonlinearity mediated via quantum-critical material fluctuations. This prior work focused on a cavity with a single-polarization mode.

    Polarization resolution offers a natural route toward making photon statistics a symmetry-sensitive probe of cavity quantum materials. In conventional Raman scattering, the dependence on incoming and outgoing photon polarizations is used to isolate distinct symmetry channels of the material response \cite{rousseau1981normal,devereaux2007inelastic}. It is therefore natural to ask whether a polarization-resolved formulation of two-photon photodetection near coincidence, $g^{(2)}(0)$, can provide a richer classification, distinguishing photon-pair scattering channels according to their symmetry and polarization structure. Such a formulation could reveal higher-order matter correlations and fluctuation-mediated photon nonlinearities that are not accessible from conventional transmission or Raman spectra alone.

    In this work we study polarization-resolved  $g^{(2)}(t\to0)$, which probes the emission of photon bunches from the cavity. We develop a theory for characterizing the full set of polarization-resolved two-photon observables. We find that polarization-selective $g^{(2)}$ measurement gives access to photon-matter dressed responses and higher-order matter correlations even at weak light-matter coupling. As a representative example, we study the Kitaev-Heisenberg spin model; a paradigmatic spin-orbit-coupled model hosting closely competing magnetic orders with distinct symmetry properties. In particular, we focus on the transition between stripy and antiferromagnetic (AFM) phases that is characterized by rotation symmetry breaking. Tuning between the two phases produces a discontinuity in the photon response at the phase boundary, reflecting the discontinuous change in material correlations at a first-order transition. We first characterize how the magnetic point groups of the two phases, $C_2$ for stripy and $C_3$ for AFM, are reflected in this polarization-dependent response. We then focus on the statistics of photon pairs transmitted with polarization orthogonal to the input polarization. Polarization-resolved photon counting can therefore simultaneously identify cavity-embedded quantum phases and phase transitions in systems with many competing magnetic orders, and characterize cavity photon-matter fluctuations.

    The rest of the paper is organized as follows: In Sec. \ref{sec:coincidence}, we introduce a theoretical framework for polarization-resolved photon coincidence statistics $g^{(2)}(t \to 0)$ in Mott insulators, and describe the effective Raman coupling between spins and optical photons. In Sec. \ref{sec:model}, we introduce the Kitaev-Heisenberg model as a paradigmatic two-dimensional magnetic model with spin-orbit interactions and categorize its coupling to light. In Sec. \ref{sec:singlePolarizationCoincidence}, we study $g^{(2)}(0)$ photon statistics near a first-order phase transition at fixed polarization in a single-mode cavity as a function of detuning and model parameters. In Sec. \ref{sec:twoPolarizationCoincidence}, we first consider the polarization-dependent photon statistics and show that they inherit the magnetic point group symmetries of the material phase. Second, we specialize to the orthogonal rotation channel, whereby $H$-polarized photon pairs are rotated to $V$-polarized emitted photons in transmission, or $L$ circularly polarized photons are rotated to $R$ circularly polarized photons. We find that this statistic is a sensitive probe of material correlations even at weak light-matter coupling and relate the polarization-selected $g^{(2)}$ response to higher-order matter fluctuations. Finally, in Sec. \ref{sec:outlook}, we summarize the results and discuss future directions.

\section{Model}

\subsection{Photon $g^{(2)}$ Measurements at Coincidence}\label{sec:coincidence}

    We first introduce a theory of coincidence measurements for polarization-resolved photons transmitted through a cavity with an embedded material, as well as its computation for cavity quantum materials. Photon coincidence is the equal time limit of second-order photon coherence measurements $g^{(2)}(t\to0)$. It can be understood as the likelihood of detecting two photons at the same time relative to the likelihood of detecting each photon individually, which, including polarization, is \cite{brown1956correlation,paul1982photon}
    \begin{align}
        g_{\mu\mu'}^{(2)}(0)=\frac{\langle\BoutXD{\mu}(0)\BoutXD{\mu'}(0)\BoutX{\mu'}(0)\BoutX{\mu}(0)\rangle}{\langle\BoutXD{\mu}(0)\BoutX{\mu}(0)\rangle\langle\BoutXD{\mu'}(0)\BoutX{\mu'}(0)\rangle},
        \label{eq:coincidenceAsBOuts}
    \end{align}
    in terms of output photon operators $\hat{b}_\mathrm{out}$, $\hat{b}_\mathrm{out}^\dag$ at a measurement time, taken to be $t=0$, with polarizations $\mu$, $\mu'$. Here, output photon operators are defined by the input-output formalism of Gardiner and Collett \cite{gardiner93,gardiner85}. Such measurements are experimentally accessible using a Hanbury Brown-Twiss setup. We consider a coherent input probe field. Values of $g^{(2)}(0)>1$ indicate photon bunching, while values of $g^{(2)}(0)<1$ indicate photon antibunching, the latter being a uniquely quantum phenomenon that cannot arise in classical electromagnetism. Changes to the photon statistics require light-matter interactions. Therefore, changes to photon statistics as they pass through the cavity can distinguish the presence of strong light-matter coupling with an embedded material from other, spurious changes to the cavity environment, at leading order in the electric field \cite{kass2024}. We further emphasize that the light-matter interactions that affect photon statistics are the same interactions that induce dark-cavity modification of material properties. In this way, the strength of bunching and antibunching observed for transmitted photons, relative to the statistics of the input field, serves as a metric for the magnitude of light-induced changes to the material.

    In particular, we focus on coincidence $g_{\mu}^{(2)}(\wIn,\nu)$ for a single output polarization $\mu=\mu'$, as a function of a coherent input field with frequency $\omega_{\textrm{in}}$ and polarization $\nu$. As detailed in Ref. \cite{kass2024}, $g_{\mu}^{(2)}(\wIn,\nu)$ takes a particularly simple form under the assumption that cavity photons interact with the material through number-conserving Raman processes \cite{sentef2020,kass2024}, in which a photon is virtually absorbed and re-emitted into the cavity while coupling to a magnetic excitation. If the optical cavity mode frequency is chosen such that it is detuned from the charge gap, photons cannot resonantly generate charge excitations. Furthermore, the light-matter coupling strength remains well below the cavity frequency for optical cavities, permitting a rotating wave approximation away from ultrastrong coupling: pair creation and annihilation processes are energetically suppressed, and we furthermore ignore multi-photon Raman processes, such as $\AD{}\AD{}\A{}\A{}$. Under these assumptions, the cavity Hamiltonian $\HamCav$ is
    \begin{align}
        \hat{H}_{\rm{cav}}=\HamMat + \AD{\rho}\A{\rho'}\left(\omega_{0}\delta^{\rho\rho'}+\DMat{\rho\rho'}\right),
        \label{eq:HamiltonianForm}
    \end{align}
    where $\HamMat$ is the dark cavity material Hamiltonian, which includes dark-cavity modification from interactions with cavity vacuum fluctuations, $\A{}$, $\AD{}$ are cavity photon operators with polarizations $\rho$, $\rho'$, and $\DMat{\rho\rho'}$ is the polarization-dependent light-matter coupling. Here we insist that the dark-cavity photon occupation vanishes, which is expected for optical cavities where the mode frequency $\omega_0$ is much larger than the strength of light-matter interactions. This permits using a Markovian input-output relation $\BoutX{\mu}(t) = \BinX{\mu}(t) - i\sqrt{\gamma} \A{\mu}(t)$ \cite{gardiner93,gardiner85} to relate the output photons $\BoutX{\mu}(t)$ to input photons $\BinX{\mu}(t)$ as a function of the cavity photon field $\A{\mu}(t)$ in Heisenberg picture \cite{gardiner85}. We note that THz cavities with non-zero dark-cavity photon occupation require using a non-Markovian scattering matrix approach \cite{talkington2026}.

    With Eq. (\ref{eq:HamiltonianForm}), we can solve the input-output relation by deriving equations of motion for $\A{\mu}(t)$ and formally integrating to get $\BoutX{\mu}(t)$ as a function of $\BinX{\mu}(t)$ \cite{kass2024}. Substituting into Eq. (\ref{eq:coincidenceAsBOuts}) one obtains
    \begin{align}
        g_{\mu}^{(2)}(\wIn,\nu) = \frac{ \sum\limits_{f} \left| \bra{f} \polTensorC_{\mu\mu} \GTen_2 \polVec_{\rm{in},\nu} \!\otimes\! \left[ \GTen_1 \polVec_{\rm{in},\nu} \!\otimes\! \ket{0} \right] \right|^2  }{ \bigg( \sum\limits_{f} \left| \bra{f} \polVecC_\mu \GTen_1 \polVec_{\rm{in},\nu} \!\otimes\! \ket{0} \right|^2\bigg)^2  }. \label{eq:g2withpolarization}
    \end{align}
    Here, the numerator gives the likelihood of observing two photons with polarization $\mu$ (i.e. $\langle \hat{b}^\dagger_\mu\hat{b}^\dagger_{\mu}\hat{b}_{\mu}\hat{b}_{\mu}\rangle$), while the denominator gives the likelihood for each photon individually (i.e. $\langle \hat{b}^\dagger_{\mu}\hat{b}_{\mu}\rangle^2$). Each piece takes the form of a scattering rate for transmitting photons through the cavity, starting from $\ket{0}$, the material ground state, and evolving to $\ket{f}$, the final material state after the photons have left the cavity. First, $\polVec_{\rm{in},\nu}$ represents a photon entering the cavity from the coherent input field with input frequency $\wIn$ and polarization vector $\polVec_{\nu}$. For a coherent drive, $g_{\mu}^{(2)}(\wIn,\nu)$ is independent of the input field amplitude, which cancels between the numerator and denominator. Instantaneously, the photon does not change the material state, but embeds it into the enlarged Hilbert space of $\ket{\rm{material}}\otimes\ket{\rm{polarization}}$. Next, the photon and the material evolve together through the resolvent  $\GTen_1$ as explained by Eq. (\ref{eq:G1Definition}) below. For the one photon likelihoods in the denominator, we then filter by the polarization vector at the detector $\polVec_{\mu}$. For the two photon likelihood in the numerator, we continue by adding a second photon from the same input field $\polVec_{\rm{in},\nu}$, embedding us into the further enlarged Hilbert space of $\ket{\rm{material}}\otimes\ket{\rm{polarization_1}}\otimes\ket{\rm{polarization_2}}$. The two photons and the material evolve together through the resolvent $\GTen_2$ as explained by Eq. (\ref{eq:G2Definition}) below. Finally, the filter for the two photons at the detector is given by $\polTensor_{\mu\mu} = 2\left(\polVec_\mu \otimes \polVec_{\mu}\right)$, where the factor of 2 handles their indistinguishability.

    As mentioned, the time evolution of the material with one and two photons in the cavity is expressed through the dressed many-body Green's resolvents $\GTen_1$ and $\GTen_2$ 
    \begin{align}
        \GTen_1 &= \frac{1}{\wIn + E_0 - \HamCavOne + i\gamma} \label{eq:G1Definition} \\ 
        \GTen_2 &= \frac{1}{2\wIn + E_0 - \HamCavTwo + 2i\gamma}, \label{eq:G2Definition}
    \end{align}
    where $E_0$ is the initial state energy, $\HamCavOne$ and $\HamCavTwo$ are the one and two photon subsectors of $\HamCav$, and $\gamma$ is the cavity linewidth. In terms of a generic polarization basis labeled $1$ and $2$, $\HamCavOne$ can be written in the enlarged $\ket{\rm{material}}\otimes\ket{\rm{polarization}}$ Hilbert space as
    \begin{align}
        \HamCavOne = \left(\HamMat+\omega_0\right)\otimes 1+\begin{pmatrix}
            \DMat{11} & \DMat{12} \\
            \DMat{21} & \DMat{22}
        \end{pmatrix},
        \label{eq:onePhotonSector}
    \end{align}
    while $\HamCavTwo$ written in the $\ket{\rm{material}}\otimes\ket{\rm{polarization_1}}\otimes\ket{\rm{polarization_2}}$ Hilbert space is
    \begin{align}
        \HamCavTwo &= \left(\HamMat+2\omega_0\right)\otimes 1\otimes 1 \notag\\ 
        &\quad+\begin{pmatrix}
            2\DMat{11} & \DMat{12} & \DMat{12} & 0 \\
            \DMat{21} & \DMat{11}\!+\DMat{22} & 0 & \DMat{12} \\
            \DMat{21} & 0 & \DMat{11}\!+\DMat{22} & \DMat{12} \\
            0 & \DMat{21} & \DMat{21} & 2\DMat{22}
        \end{pmatrix}\!.
    \end{align}
    The four-dimensional two-photon polarization space carries a permutation symmetry between the pair of polarizations; the output polarization tensor ($\polTensor_{\mu\mu}$) subsequently projects on the physical polarization state space of three distinguishable polarizations $HH$, $VV$, $HV = VH$ (or, equivalently, $LL$, $RR$, $LR = RL$) of two photons in the single-mode cavity.

    Eqs. (\ref{eq:G1Definition}) and (\ref{eq:G2Definition}) express that after one (two) photons enter the cavity, the cavity evolves according to $\HamCav$ in the one (two) photon sector. During this evolution, the photons and material become entangled, and material properties can affect the photon statistics observed in the bunching and antibunching of transmitted photons. In analogy to the process where individual emitters result in a photon blockade effect \cite{goto2019figure,reiserer2015}, this process involving many-body quantum materials when leading to antibunching is referred to as a many-body photon blockade.

\subsection{Cavity-Embedded Kitaev-Heisenberg Magnets}\label{sec:model}

    \begin{figure*}
        \centering
        \includegraphics[width=1.0\textwidth]{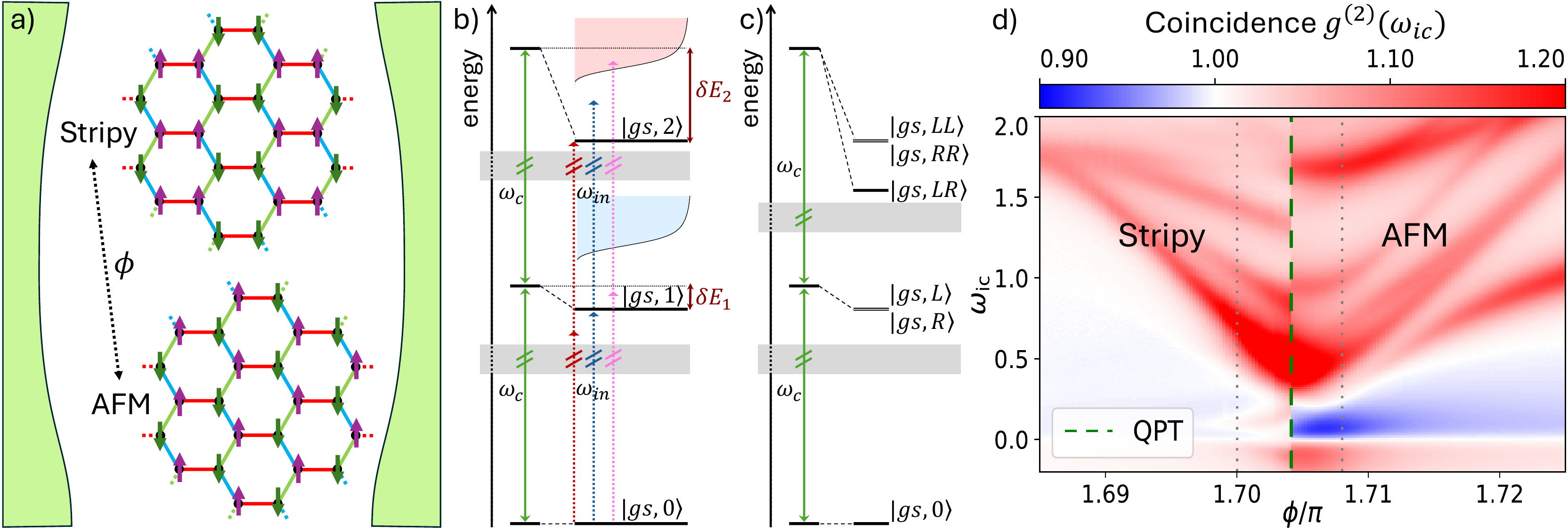}
        \caption{\textbf{Many-Body Photon Blockade.} \textbf{(a)} The cavity-embedded Kitaev-Heisenberg spin model is a paradigmatic model hosting a rich set of competing magnetic phases that can hybridize with cavity photons. Classical ordering patterns of spins are shown for stripy and N\'{e}el-antiferromagnet (AFM) phases, tuned via a mixing angle $\phi$ for Kitaev and Heisenberg interactions. \textbf{(b)} Energy levels of the $n$-photon sector ground states $\ket{gs,n}$, with a single cavity mode. Without light-matter coupling, these states are equally separated by the cavity mode frequency $\omega_c$ (green arrows). With light-matter coupling, the photons and material hybridize, causing energy shifts $\delta E_n\propto n^2$ that are non-linear in the photon number $n$. This results in distinct one- and two-photon resonances, leading to antibunching (bunching) of photons transmitted through the cavity. Red, blue, and pink arrows demonstrate transitions from bunching to antibunching and back to bunching as the input frequency $\omega_{\rm{in}}$ increases, with the highest frequency being two photon resonant with excited states (blue and pink regions). \textbf{(c)} With polarization, $\ket{gs,n}$ splits into $n+1$ states with different numbers of left- ($L$) vs right- ($R$) circularly polarized photons in the cavity. Note that $\ket{gs,LR}$ splits in energy from $\ket{gs,LL}$ and $\ket{gs,RR}$. \textbf{(d)} Single-polarization coincidence, aligned with the $z$ bonds, as a function of input frequency detuning $\wIC$ as $\phi$ is tuned across the stripy-to-AFM quantum phase transition (QPT) with $\gamma=0.1$, $g=0.3$, $\omega_{\textrm{0}}=0.6$, and $B=0.001$. The discontinuities in bunching and antibunching are indicative of a first-order QPT. The gray dotted lines represent stripy and AFM points studied in more detail in Sec. \ref{sec:twoPolarizationCoincidence}. 
        }
        \label{fig:model_and_phase}
    \end{figure*}

    To illustrate how polarization-resolved coincidence measurements can diagnose phases of matter in a cavity quantum material, we focus on the Kitaev-Heisenberg model as a simple, but paradigmatic example of closely competing magnetic orders. This model has a rich and well-studied phase diagram \cite{Jackeli2009,chaloupka2013,rau2014,Winter2016,Winter2017,gotfryd2017,Winter2017B}. We want to focus on a transition between phases with different rotational symmetries, and therefore focus on the first-order stripy-to-AFM quantum phase transition (QPT). In the thermodynamic limit, the stripy phase has $C_2$ rotational symmetry compared to the $C_3$ rotational symmetry of the AFM phase. Fig. \ref{fig:model_and_phase}(a) shows the classical spin pattern for these phases on the 24-site honeycomb cluster with periodic boundary conditions used. In finite-sized systems, however, the three stripy orientations mix, such that the stripy ground state does not break $C_3$ symmetry. Therefore, a biasing mechanism is applied to pick out one orientation as would naturally occur in the thermodynamic limit. Similarly, the finite sized ground state of the AFM phase forms from a superposition of states rotated by 180 degrees, such that it has $C_6$ symmetry. We do not apply biasing to the AFM phase, however, since this does not affect the symmetries seen by linearly polarized photons. Details of the phase transition and the biasing process are given in Appendix \ref{app:kitaevMaterialPhases}.

    The Kitaev-Heisenberg spin Hamiltonian with an applied magnetic field is
    \begin{align}
        \HamMat &= \sum_{\langle ij\rangle}\left(J\hat{S}_i\hat{S}_j +K\hat{S}_i^\gamma\hat{S}_j^\gamma\right) + B \sum_i \left( \hat{S}_i^x + \hat{S}_i^y + \hat{S}_i^z \right). \label{eq:khHam}
    \end{align}
    Here, $\hat{S}_i$ is the spin at site $i$. $\langle ij\rangle$ are nearest neighbor bonds, for which $\gamma$ is a bond-dependent Kitaev spin direction with $\gamma\in\left\{x,y,z\right\}$, as depicted in blue, green, and red respectively in Fig. \ref{fig:model_and_phase}(a). $J$ and $K$ are nearest-neighbor Heisenberg and Kitaev coupling strengths. $B$ is an applied magnetic field strength, which we choose to lie in the [111] direction in order to preserve $S_6$ improper rotational symmetry of the system, which causes the spins to permute into each other along with lattice rotations as $\hat{S}^x\to\hat{S}^y\to\hat{S}^z\to\hat{S}^x$. We follow the convention to normalize $J$ and $K$, such that $J^2+K^2=1$, and parametrize with $\phi$ as $J=\cos\phi$ and $K=\sin\phi$.

    To lowest order, photons couple to a spin Hamiltonian through spin exchange processes \cite{sentef2020,sriram2022light,vinas2023controlling}
    \begin{align}
        \DMat{\mu\nu}=\sum_{\langle ij\rangle}\sum_{\sigma_1,\sigma_2}D^{\mu\nu}_{ij,\sigma_1\sigma_2}\hat{S}_i^{\sigma_1}\hat{S}_j^{\sigma_2}.
    \end{align}
    The allowed terms and their relative strengths depend on the material parameters $J$ and $K$, as well as the polarization sector ($\mu\nu$) and the bond type for nearest neighbors $\langle ij\rangle$ (Kitaev $x$, $y$, or $z$). To estimate reasonable light-matter coupling strengths for Kitaev materials, we follow the approach in Ref. \cite{sriram2022light,vinas2023controlling}. We focus on a single Kitaev $z$ bond, starting from the octahedrally arranged $p$- and $d$-orbitals that lead to Kitaev interactions along it \cite{Jackeli2009}. We use a Peierls' substitution to couple parallel and perpendicularly polarized cavity photons to each hopping term. We down-fold the full Hamiltonian to the spin sector and separate per photon sector to get this bond's contribution to $\HamMat$ and $\HamCavOne$. Then, we invert Eq. (\ref{eq:onePhotonSector}) to get $\DMat{11}$, $\DMat{12}$, $\DMat{21}$, and $\DMat{22}$ for the bond. For the Kitaev-Heisenberg model, we find that the contribution to $\DMat{\mu\nu}$ from a Kitaev $z$ bond is
    \begin{align}
        \DMat{\mu\nu}_{z\text{-bond}}=\!\sum_{\sigma}D^{\mu\nu}_{z,\sigma\sigma}\hat{S}_i^\sigma\hat{S}_j^\sigma+D^{\mu\nu}_{z,xy}\!\left(\hat{S}_i^x\hat{S}_j^y+\hat{S}_i^y\hat{S}_j^x\right)
    \end{align}
    Finally, we rotate the resulting spin coupling parameters by $2\pi/3$ in the polarization basis to obtain couplings for the Kitaev $x$ and $y$ bonds. This approach requires starting from a full set of \textit{electronic} material parameters. As there is not a unique mapping from $J$, $K$ to these material parameters, we use the values taken from \cite{vinas2023controlling} for \ce{\alpha-RuCl3} and fit the others to $J$ and $K$ \seeAppendixX{\ref{app:lightMatterCouplingFitting}}.

    The values of $J$, $K$, and $\phi$ in $\HamMat$ are the dark-cavity modified material parameters. Their origin is rooted in the same processes that give rise to the photon-matter coupling terms $\hat{D}^{\mu\nu}$---both photon vacuum fluctuations and real photons alter the superexchange processes that couple to spin. In the presence of an optical cavity mode that is red-detuned from charge excitations, integrating out charge generically leads to effective magnetic exchange couplings that are operator-valued functions $\hat{J}$, $\hat{K}$ of the cavity photon field \cite{sentef2020,vinas2023controlling}. For instance, expanded to leading order in light-matter coupling, one finds $\hat{J} = J(g) + \hat{D}^{\mu\nu}(g) \AD{\mu} \A{\nu} + \cdots$ where $g$ is the light-matter coupling strength and $J(g)$ is the vacuum-fluctuation-dressed exchange coupling. Photon coincidence measurements [Eq. (\ref{eq:g2withpolarization})] probe the ground state of the material including such cavity-induced changes, by measuring how this ground state changes under the addition of photons. The cavity-modified state can then be compared to similar measurements of the ground state outside the cavity \cite{nambiar2025diagnosing}. In this way, photon coincidence measurements can probe material changes and phase transitions and the strength of the $g^{(2)}$ response can help isolate light-induced material changes from the impact of other, spurious changes due to the cavity environment. Since both the light-matter coupling strengths $D^{\mu\nu}_{ij,\sigma_1\sigma_2}$ and the dark-cavity modification of $J$ and $K$ enter at second-order in the light-matter coupling strength $g$ for spin models \cite{sentef2020}, the magnitude of the bunching/antibunching response will directly correspond to the magnitude of light-matter-coupling-induced material changes.

\section{Results}

\subsection{Photon $g^{(2)}$ at Coincidence From Single-Polarization Cavities}
\label{sec:singlePolarizationCoincidence}

    To ground the polarization-resolved response, we first look at coincidence measurements across the stripy-to-AFM transition in a cavity that only allows a single photon polarization mode aligned with the Kitaev $z$ bonds. We study the 24-site cluster with periodic boundary conditions shown in Fig. \ref{fig:model_and_phase}(a), chosen as the largest tractable system size that preserves six-fold rotational symmetry. We limit to the zero momentum sector, since $q\approx0$ photons cannot noticeably change the momentum in the material. We use Lanczos exact diagonalization to get the material ground state $\ket{0}$ and a biconjugate gradient stabilized linear solver to compute products of resolvent operators $\GTen_1$, $\GTen_2$ and many-body states.

    Fig. \ref{fig:model_and_phase}(d) shows the resulting coincidence spectrum $g^{(2)}(\wIC)$ versus the $J$-$K$ mixing angle $\phi$ with model parameters specified in the caption. As previously discussed, $\phi$ includes dark-cavity material changes, and so $\phi$ could be tuned either by changing the light-matter coupling strength or by changing the intrinsic material parameters. The input frequency detuning $\wIC=\wIn-\omega_{\textrm{c}}$ subtracts the elementary cavity one-photon resonance $\omega_{\textrm{c}}$, which can be obtained experimentally via transmission and can be interpreted as a shifted bare cavity resonance due to changes in the intra-cavity dielectric constant from the material. For each $\phi$, we fit $\omega_c=\omega_{\rm{max},\phi}$ to the frequency that maximizes single photon transmission. This shifts the reference point for input frequencies in order to center important spectral features around 0 \seeAppendixX{\ref{app:cavityModeFrequency}}. The QPT at $\phi\approx1.704\pi$ was identified through the extremum of the second derivative of ground state energies and is marked by the dashed green line.

    We now relate features of the single polarization coincidence spectrum to material excitations in the cavity. For example, a region of antibunching appears to sit in the excitation gap, while local maxima of bunching at larger input frequencies appear to track Raman active excitations shown in Fig. \ref{fig:phase_info}(c). To understand these features, we consider the energy level spectrum depicted in Fig. \ref{fig:model_and_phase}(b) which shows matter excitations dressed by $n$ photons injected into the cavity. For perturbative light-matter coupling strengths $D\ll1$, the ground state $\ket{gs,n}$ of the $n$-photon sector shifts downward by $\delta E_n\propto n^2$ relative to the individual material and photon energies. This comes directly from the second-order perturbative expansion
    \begin{align}
        \delta E_n= -\sum_{i>0}\frac{\bra{0}n\hat{D}\ket{i}\bra{i}n\hat{D}\ket{0}}{E_i-E_0} + \mathcal{O}(D^3),
    \end{align}
    where $E_i$ are the energies for eigenstates $\ket{i}$ of the dark cavity Hamiltonian $\HamMat$, and where we have dropped the polarization dependence from $\hat{D}$ (as we consider a cavity with a fixed photon polarization). We also exclude linear shifts in the $n$-photon ground state energy, which are already included in the definition of $\omega_c$. As demonstrated by the red, blue, and pink arrows in Fig. \ref{fig:model_and_phase}(b), different input frequencies will be one or two photon resonant. Sweeping across frequencies from red to blue detuning with respect to the bare cavity resonance, first the red arrow is two photon resonant with $\ket{gs,2}$, causing bunching at the lowest frequencies. Next, the blue arrow becomes one photon resonant with $\ket{gs,1}$, causing antibunching at intermediate frequencies. We note that this one photon resonance also determines our choice of reference frequency $\omega_{\rm{c}}$, such that the transition from bunching to antibunching will occur at input frequency detunings $\wIC\lesssim0$. Finally, the pink arrow becomes two photon resonant with excitations of the two photon sector (pink region), giving bunching again. As long as there is an energy gap $\Delta$ to the lowest Raman active excited state, this latter transition from antibunching to bunching is expected to occur around $\wIC\gtrsim\Delta/2$ where the two-photon resonance invokes a matter excitation.

    To be more precise about connecting the patterns of bunching and antibunching to material correlation functions, we derive a perturbative expansion for $g^{(2)}(\wIC)$ in powers of $D$. We first fix $\omega_c = \omega_0 + \bra{0} \hat{D} \ket{0}$, shifted by the ground state expectation value of $\hat{D}$. This permits redefining the resolvents ($\GTen_1$, $\GTen_2$) by shifting $\hat{D} \to \delta\hat{D} = \hat{D} - \bra{0} \hat{D} \ket{0}$, the \textit{fluctuations} about the ground state expectation value. For conciseness, we write $\delta\hat{D} \to \hat{D}$ after this redefinition. We now expand these resolvent operators in Eq. (\ref{eq:g2withpolarization}) as a Neumann series in terms of $\hat{D}$, insert complete sets of eigenstates, and simplify in terms of matrix elements $D_{ij}=\bra{i}\hat{D}\ket{j}$ and energy denominators \seeAppendixX{\ref{app:perturbativeExpansion}}. We get to second order in $D$
    \begin{align}
        g^{(2)}(\wIC) = 1 + \!\sum_i \frac{4\wIC\left(\wIC-\Delta E_i/2\right)D_{0i}D_{i0}}{\left(\wIC^2+\gamma^2\right)\left(\left(\wIC\!-\Delta E_i\right)^2\!+\gamma^2\right)},
        \label{eq:singlePolarizationCoincidenceExpansion}
    \end{align}
    where $\Delta E_i=E_i-E_0$ is the excitation energy for the corresponding eigenstate $\ket{i}$ of $\HamMat$. This can be usefully written in terms of the two-point dynamic material correlation functions as
    \begin{align}
        g^{(2)}(\wIC) = 1 + \int d\omega\,\frac{4\wIC\left(\wIC-\omega/2\right)\chi(\omega)}{\left(\wIC^2+\gamma^2\right)\left(\left(\wIC\!-\omega\right)^2\!+\gamma^2\right)}.
        \label{eq:singlePolarizationCoincidenceExpansionCorrelators}
    \end{align}
    where
    \begin{align}
        \chi(\omega)=\int dt\,e^{i\omega t}\left< \hat{D}(t)\hat{D}(0)\right>
    \end{align}
    is the material's Raman dynamical structure factor \cite{devereaux2007inelastic}.

    The constant of 1 comes from forward scattering, the elastic tunneling of photons through the cavity without material interactions, which leaves their statistics unchanged. The sign of the second order correction is determined entirely by $\wIC(\wIC-\Delta E_i/2)$, since all the other terms are positive. Therefore, each Raman-active energy level $E_i$ contributes negative corrections to $g^{(2)}$ (towards antibunching) when $0<\wIC<\Delta E_i/2$ and positive corrections (towards bunching) when $\wIC<0$ or $\wIC>\Delta E_i/2$. This confirms the intuition from one- and two-photon resonances described above, ensuring bunching for $\wIC<0$ and antibunching for detunings inside the energy gap $0<\wIC<\Delta/2$. Also, for the upper range of bunching ($\wIC>\Delta E_i/2$), the peak of maximal bunching in (d) occurs near the resonance at $\wIC=\Delta E_i$. In addition, this perturbative approach makes clear that the magnitude of bunching and antibunching scales as $\mathcal{O}(ND^2)$ in the perturbative regime, from which one can infer the magnitude of dark-cavity-induced changes to the spin coupling strengths $J$ and $K$.

    For continuous QPTs, the input-photon detuning region that leads to antibunching disappears (up to linewidth $\gamma$ resolution) as Raman-active quantum-critical fluctuations become gapless \cite{kass2024}. In contrast, the gap to Raman-active excitations remains finite at the first-order transition between AFM and stripy magnetic order, in the absence of domain formation. The competing ground states that cross at the transition have macroscopically distinct spin ordering patterns and cannot be coupled by local Raman interactions $\hat{D}$. Therefore, a finite detuning window of antibunching persists in coincidence measurements at the transition. Instead, first-order QPTs are identified by a discontinuity in the response, originating from the abrupt reconfiguration of the ground state and the resulting shift in Raman active excitations. The discontinuity shows up clearly in Fig. \ref{fig:model_and_phase}(d) as a shift in the bunching peaks and as a jump in the antibunching strength, both of which are absent in a continuous QPT.

\subsection{Polarization-Resolved Photon $g^{(2)}$ at Coincidence}
\label{sec:twoPolarizationCoincidence}

    Next, we extend our investigations to polarization-resolved coincidence measurements $g_{\mu}^{(2)}(\wIC,\nu)$ from a cavity with two degenerate polarization modes. Specifically, we consider measurements of same-polarization-$\mu$ output-photon pairs in response to a polarization-$\nu$ input field. The relation between input and output polarizations provides an extra degree of freedom for probing how a material interacts with light in different symmetry channels. In Sec. \ref{sec:symmetryBreaking}, we first study how $g_{\mu}^{(2)}(\wIC,\nu)$ measurements reflect symmetries of the cavity-embedded material as we change $\nu$ and $\mu$. For a given input polarization $\nu$, a striking difference emerges between measuring same-polarization $\mu=\nu$ or orthogonal (``rotated'') polarization $\mu \perp \nu$ pair coincidences $g^{(2)}(0)$. When $\mu=\nu=1$, the second order perturbative expansion for the unrotated response has the exact same form as Eq. (\ref{eq:singlePolarizationCoincidenceExpansionCorrelators}), except with $\chi(\omega)$ replaced by a polarization-dependent material correlator
    \begin{align}
        g_u^{(2)}(\wIC) &= g_1^{(2)}(\wIC,1)\notag\\
        &=1 + \int d\omega\,\frac{4\wIC\left(\wIC-\omega/2\right)\chi^{11}(\omega)}{\left(\wIC^2+\gamma^2\right)\left(\left(\wIC-\omega\right)^2+\gamma^2\right)}\label{eq:unrotatedCoincidencePerturbativeCorrelation}\\
        \chi^{11}(\omega)&=\int dt\,e^{i\omega t}\left< \hat{D}^{11}(t)\hat{D}^{11}(0)\right>,
    \end{align}
    where $\omega_c = \omega_0 + \bra{0} \DMat{11} \ket{0}$. The same-polarization coincidence response $g_1^{(2)}(\wIC,1)$ through a cavity with two polarizations only differs from the single polarization $g_1^{(2)}(\wIC,1)$ by corrections to higher order in light-matter coupling $D$ \seeAppendixX{\ref{app:comparisonOfOneAndTwo}}. The rotated response with $\mu=2$ orthogonal to $\nu=1$, however, takes a very different form. We derive a perturbative expansion for $g_r^{(2)}(\wIC) = g_2^{(2)}(\wIC,1)$ and study its relationship to material properties in Sec. \ref{sec:rotatedResponse}.

\subsubsection{Symmetry Breaking in Polarization-Resolved Photon $g^{(2)}$ at Coincidence}
\label{sec:symmetryBreaking}

    \begin{figure}
        \centering
        \includegraphics[width=1.0\columnwidth]{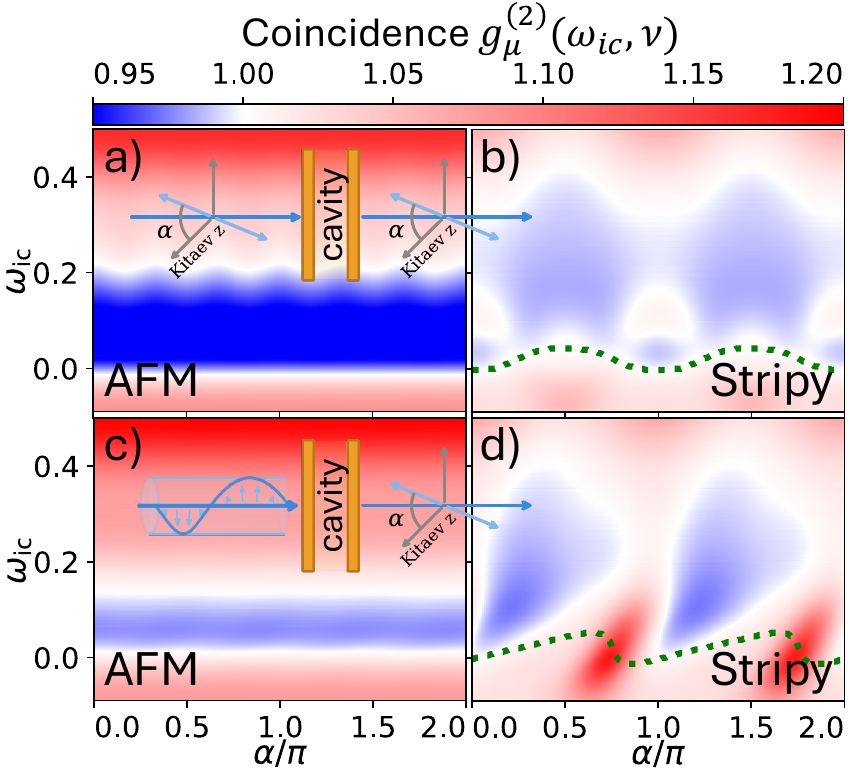}
        \caption{\textbf{Polarization-Dependent Coincidence.} Coincidence response in a cavity with two polarization modes as a function of the output photon linear polarization angle $\alpha$ in the plane of the material relative to the Kitaev $z$ bonds. The top row shows the response to linearly polarized input photons with the same angle $\alpha$ while the bottom row uses right circularly polarized input photons, as depicted in the overlaid diagrams. \textbf{(a)} and \textbf{(c)} Coincidence for the AFM phase at $\phi=1.708\pi$. \textbf{(b)} and \textbf{(d)} Coincidence for the stripy phase at $\phi=1.7\pi$, biased along Kitaev $z$ bonds. These correspond to the gray dotted lines in Fig. \ref{fig:model_and_phase}(d). As in Fig. \ref{fig:model_and_phase}(d), $\gamma=0.1$, $g=0.3$, $\omega_{\textrm{0}}=0.6$, and $B=0.001$. Green dotted lines in (b) and (d) track the maximal one photon transmission vs polarization angle. Note that the rotational and mirror symmetries of the response reflect the symmetries of the respective phases. }
        \label{fig:two_polarization_results}
    \end{figure}

    \begin{figure*}
        \centering
        \includegraphics[width=1.0\textwidth]{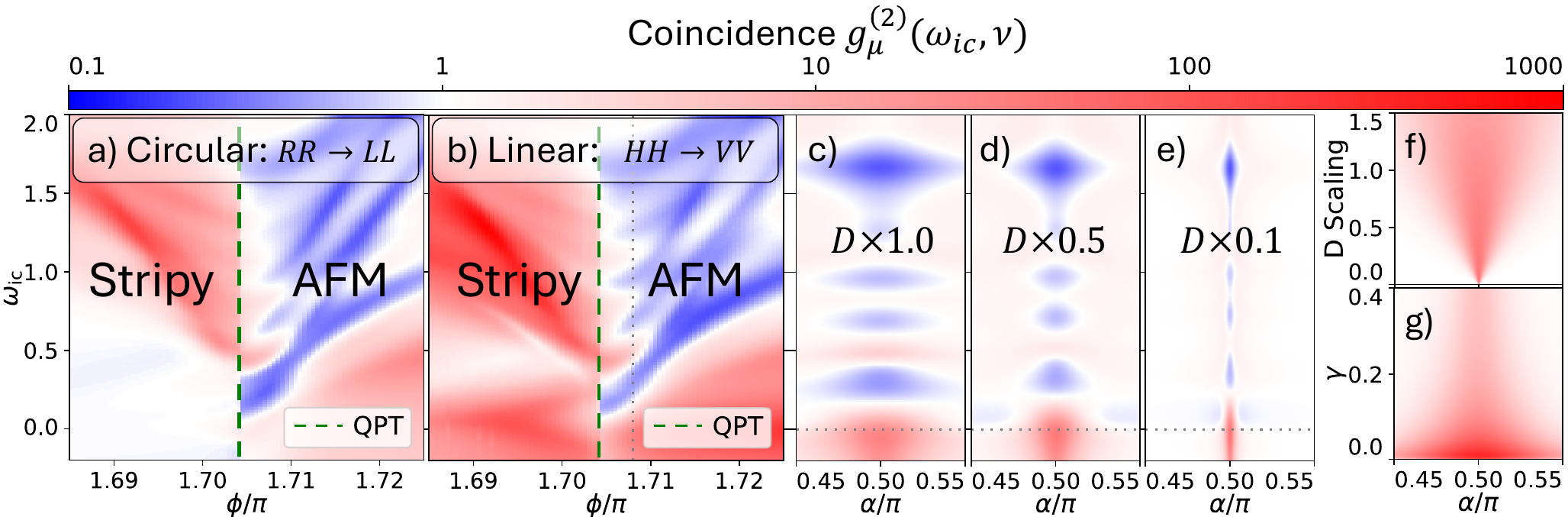}
        \caption{\textbf{Polarization Rotation.} Coincidence response across the stripy-to-AFM phase transition of Fig. \ref{fig:model_and_phase}(d) for photons rotated from the input polarization. \textbf{(a)} Circularly polarized photons input with right ($R$) and output with left ($L$) circular polarization. \textbf{(b)} Linearly polarized photons input aligned with Kitaev $z$ bonds ($H$) and output perpendicular to them ($V$). Coincidence vs linearly polarized output rotation angle $\alpha$ relative to input light along the $z$ bond for the AFM point at $\phi=1.708\pi$ corresponding to the gray dotted line in (b), as the light-matter coupling strength $D$ is scaled by \textbf{(c)} 1.0, \textbf{(d)} 0.5, and \textbf{(e)} 0.1 relative to its value in Fig. \ref{fig:model_and_phase} and in Sec. \ref{sec:model}. \textbf{(f)} Coincidence vs $\alpha$ and scaling of the light-matter coupling strength $D$ for input frequency detuning $\wIC=0$, corresponding to the gray dotted lines in (c)-(e). \textbf{(g)} Similar for coincidence vs $\alpha$ and the linewidth $\gamma$. Note that the bunching and antibunching remain strong at exactly $\alpha=\pi/2$, even as $D\to0$. In all plots, $\gamma=0.1$, $g=0.3$, $\omega_{\textrm{0}}=0.6$, and $B=0.001$ as before (except where otherwise specified). }
        \label{fig:polarization_rotation}
    \end{figure*}

    First, we consider how polarization-resolved coincidence measurements reflect symmetries of the material. Eq.  (\ref{eq:unrotatedCoincidencePerturbativeCorrelation}) shows that the unrotated polarization-resolved response is a function of $\chi^{11}(\omega)$, a polarization-dependent material correlation function that must reflect rotation and mirror symmetries of the material. Therefore, we expect to see such symmetries reflected in the unrotated response as we change the incoming photon polarization. Figs. \ref{fig:two_polarization_results}(a) and (b) show $g_u^{(2)}(\wIC)=g_\alpha^{(2)}(\wIC,\alpha)$ for linearly polarized incoming and outgoing photons as we rotate their angle $\alpha$ relative to the Kitaev $z$ bonds, as depicted in the overlaid diagram. Panel (a) is the response for the AFM phase at $\phi=1.708\pi$, while (b) is the response for the stripy phase at $\phi=1.7\pi$, corresponding to the two gray dotted lines in Fig. \ref{fig:model_and_phase}(d). For each, we fit the reference $\omega_c=\omega_{\rm{max},\alpha=0}$ to the frequency that maximizes single photon transmission of linearly polarized photons aligned with the Kitaev $z$ bonds \seeAppendixX{\ref{app:cavityModeFrequency}}.

    As expected from the reduction of rotational symmetry in the material across the QPT, the two states show dramatically different polarization-resolved responses. The AFM phase in Fig. \ref{fig:two_polarization_results}(a) shows $C_6$ rotational symmetry, as well as mirror lines parallel and perpendicular to each bond of the honeycomb lattice. The stripy phase in (b) only has $C_2$ rotational symmetry and mirror lines parallel and perpendicular to the Kitaev $z$ bonds. The $C_2$ rotational symmetry of the stripy phase also introduces anisotropy to the cavity, lifting the degeneracy and splitting the frequencies of the two polarization modes. Therefore, $\omega_{\rm{max},\alpha}$ becomes a function of $\alpha$, which we plot as the dotted green line in (b). The transition from antibunching to bunching in (b) tracks this polarization-dependent cavity mode frequency quite well. Finally, the two regions of bunching in the stripy phase around $\alpha\in\left\{0,\pi\right\}$ are due to finite size splitting of the ground state appearing as low energy excitations, and therefore would not appear in the thermodynamic limit. This same effect also appears as a trail of bunching extending across the QPT into the stripy phase of Fig. \ref{fig:model_and_phase}(d). Comparing against the Raman active excitations in Fig. \ref{fig:phase_info}(c), we see that this trail follows states that gradually merge into the ground state on the stripy side, which would occur exactly at the QPT in the thermodynamic limit.

    In Figs. \ref{fig:two_polarization_results}(c) and (d), we consider $g_\alpha^{(2)}(\wIC,R)$ with incoming right circularly polarized photons ($R$) observed at the same outgoing linear polarization angle $\alpha$ as before, as depicted in the overlaid diagram. We also keep the same reference cavity mode $\omega_{\rm{c}}=\omega_{\rm{max,0}}$ as in (a) and (b). This coincidence response is a mix between rotated and unrotated responses, so need not behave exactly as Eq.  (\ref{eq:unrotatedCoincidencePerturbativeCorrelation}). In the stripy phase (d), the response to circularly polarized input still has $C_2$ rotational symmetry, but the photons break the mirror symmetries. Also, the maximum one photon transmission (green dotted line) no longer tracks the transition from bunching to antibunching, due to the rotated coincidence response, as we will see in Sec. \ref{sec:rotatedResponse}. In the AFM phase (c), the $C_6$ rotational symmetry is enhanced to continuous rotational symmetry for circularly polarized light. This is generic for any phase with $C_3$ symmetry, which can be explained through a group theoretic argument based on the phase picked up under rotations when the response tensor is written in the circularly polarized basis \seeAppendixX{\ref{app:continuousRotationalSymmetry}}.

\subsubsection{Rotated Response in Polarization-Resolved Photon $g^{(2)}$ at Coincidence}
\label{sec:rotatedResponse}

    We now show that polarization filtering provides a direct way to suppress the background of forward-scattered photons and isolate nonlinear material correlations. We consider the coincidence response in which two $H$-polarized input photons are converted into two $V$-polarized output photons, or equivalently, two right-circular ($R$) polarized photons are converted into two left-circular ($L$) polarized output photons. Because such a two-photon polarization rotation cannot arise from photons simply passing through the cavity, both detected photons must scatter from the material. This post-selection therefore singles out higher-order matter correlation functions while eliminating the dominant unrotated transmission background.

    Figs. \ref{fig:polarization_rotation}(a) and (b) show such rotated coincidence spectra vs input frequency detuning $\wIC$ across the stripy-to-AFM phase transition, using the same parameters as in Fig. \ref{fig:model_and_phase}(d). Panel (a) focuses on transmitted photons that are rotated from right- ($R$) to left- ($L$) circular polarization, while (b) shows $g^{(2)}(0)$ for linearly-polarized photons which are injected into the cavity with polarization aligned with the Kitaev $z$ bonds and are emitted perpendicular to them. In the stripy phase, there is very little antibunching. In (a), we see bunching at detunings corresponding to material excitations, while in (b) we see even stronger bunching that extends to most input frequencies, including those that gave antibunching in the unrotated response. In the AFM phase, both (a) and (b) show strong bunching at detunings inside the material excitation gap, and strong antibunching when resonant with material excitations. The strength of the bunching and antibunching is dramatically enhanced for the rotated photon coincidence, and therefore we show Fig. \ref{fig:polarization_rotation} on a log color scale.

    In contrast to the transmission of pairs of photons with unrotated polarization, the rotated coincidence $g_r^{(2)}(\wIC)=g_{\mu}^{(2)}(\wIC,\nu)$ (where $\mu \perp \nu$ are orthogonal polarizations) is not captured by simple resonance matching conditions between the input field and one- or two-photon cavity resonances. To see this, we consider the case where photons are rotated from $R$ to $L$ polarization. Given angular momentum conservation, the $n$-photon sector energy levels naturally split into eigenstates with $n_L$ photons of predominantly $L$ character and $n_R=n-n_L$ photons of predominantly $R$ character, as in Fig. \ref{fig:model_and_phase}(c). Injecting input photons populates the $\ket{gs,R}$ and $\ket{gs,RR}$ one- and two-photon states, but emission of these photon polarizations is discarded by the output filter. Instead, we need to perform a detailed analysis of the mixing of $L$ and $R$ character of the one and two photon sector eigenstates, caused by light-matter coupling $\DMat{LR}$ \seeAppendixX{\ref{app:rotatedDetails}}.

    To understand the transmission of pairs of rotated photons, it is much more instructive to use a perturbative expansion in $D$, thereby translating the polarization-rotated $g_r^{(2)}(\wIC)$ response into matter correlation functions. We perform the same procedure that led to Eq. (\ref{eq:unrotatedCoincidencePerturbativeCorrelation}), but only keep terms that rotate both photons \seeAppendixX{\ref{app:perturbativeExpansion}}. This requires going to fourth order in $D$ for two photon transmission, and thus has sensitivity to a four-point material correlation function. To lowest order in $D$, the rotated response is
\begin{widetext}
    \begin{align}
        g_r^{(2)}(\wIC)=g^{(2)}_2(\omega_{\rm{ic}},1) = \dfrac{\displaystyle\sum_{jkl}\dfrac{D^{12}_{0j}D^{12}_{jk}D^{21}_{kl}D^{21}_{l0}}{\left(\wIC-\Delta E_j-i\gamma\right)\left(\left(\wIC-\Delta E_k/2\right)^2+\gamma^2\right)^2\left(\wIC-\Delta E_l+i\gamma\right)}}{\left(\displaystyle\sum_i\dfrac{D^{12}_{0i}D^{21}_{i0}}{\left(\wIC-\Delta E_i\right)^2+\gamma^2}\right)^2}+\mathcal{O}(D),
        \label{eq:rotatedCoincidencePerturbative}
    \end{align}
    which can be written in terms of the two- and four-point dynamic material correlation functions  as 
    \begin{align}
        g_r^{(2)}(\wIC)=g^{(2)}_2(\omega_{\rm{ic}},1) = \dfrac{\displaystyle\int d\omega_1d\omega_2d\omega_3\,\dfrac{K_{11}^{22}(\omega_1,\omega_2,\omega_3)}{\left(\wIC-\omega_3-i\gamma\right)\left(\left(\wIC-\omega_2/2\right)^2+\gamma^2\right)^2\left(\wIC-\omega_1+i\gamma\right)}}{\left(\displaystyle\int d\omega\,\dfrac{\chi_1^2(\omega)}{\left(\wIC-\omega\right)^2+\gamma^2}\right)^2}+\mathcal{O}(D),
        \label{eq:rotatedCoincidencePerturbativeCorrelation}
    \end{align}
    where
    \begin{align}
        \chi_1^2(\omega)&=\int dt\,e^{i\omega t}\bra{0}\DMat{12}(t)\DMat{21}(0)\ket{0}\notag\\
        K_{11}^{22}(\omega_1,\omega_2,\omega_3)&=\int dt_1dt_2dt_3\,e^{i\omega_1 t_1}e^{i\omega_2 t_2}e^{i\omega_3 t_3}\bra{0}\DMat{12}(t_1+t_2+t_3)\DMat{12}(t_1+t_2)\DMat{21}(t_1)\DMat{21}(0)\ket{0}
    \end{align}
\end{widetext}
    The exact values for these equations depend on a subtle balance of matrix elements and energy denominators. However, there are a few general statements that we can make. First, resonances still occur around Raman active excitation energies $\wIC=\Delta E_m$. If the lowest-lying Raman-active excitation is energetically isolated, this resonance shows up in the unrotated response as a local maximum in bunching. In the rotated response, however, the resonance shows up as a local minimum, as we get a fourth-order pole in the denominator of Eq. (\ref{eq:rotatedCoincidencePerturbative}), but only a second-order pole in the numerator. Physically, the one photon resonance is stronger when the photons interact with the material individually than when their interactions overlap. This stands in contrast to the same-polarization case where photons transmitted in the absence of material scattering contribute $g^{(2)}=1$ with material scatterings causing modifications of order $D^2/\gamma^2$; here the only transmitted photons are those that are rotated by the material and $g^{(2)}$ is entirely dependent on the order $D^4/\gamma^4$ terms. The resulting local minimum gives antibunching for $\gamma\to0$, but could also appear as a region of decreased bunching for finite linewidths, depending on the matrix elements involved. There are also resonances at $\wIC=\Delta E_m/2$ in the numerator of Eq. (\ref{eq:rotatedCoincidencePerturbative}), which therefore show up in the rotated response as a local maximum in bunching. This peak in bunching corresponds to the transition from bunching to antibunching in the unrotated response. Also, while all of these features entered as $D^2$ corrections for the unrotated response, they are independent of $D$ to leading order for the rotated response, and therefore can be much stronger. This is a consequence of the filtering of the background of forward scattered photons, since each rotated photon must have interacted with the material.

    Figs. \ref{fig:polarization_rotation}(c)-(e) demonstrate how the linear-to-linear rotated polarization response changes as we scale $D$ for the AFM phase $\phi=1.708\pi$, corresponding to the gray dotted line in (b). We show $g^{(2)}_\alpha(\omega_{\rm{ic}},0)$ as a function of the rotation angle $\alpha$ for linearly polarized output photons relative to input photons linearly polarized along the Kitaev $z$ bonds. While $g_r^{(2)}(\wIC)$ corresponds to exactly $\alpha=\pi/2$, we see that a similar response can be observed over a range of angles $\alpha\approx\pi/2$. As $\alpha$ deviates from $\pi/2$, the detectors now also pick up unrotated photons (with the same polarization as the input field). However, these photon events do not require scattering off the material and will be dominated by forward scattering for weak light-matter coupling that pushes $g^{(2)}$ towards one. The range of $\alpha$ angles of the observed deviation of $g^{(2)}$ from unity therefore depends on the relative strength of the rotated signal relative to the number of unrotated photons that get forward scattered. If too many forward scattered photons are picked up at a given intermediate angle, the rotated response will be drowned out. As $D$ is decreased, the rotated-photon contribution becomes smaller relative to the forward scattering and the signal (deviations of $g^{(2)}$ from one) is lost at angles $\alpha$ away from $\pi/2$. At exactly $\alpha=\pi/2$, all forward scattering contributions are removed from the measured signal by polarization selection at the detector, to single out the two-photon-scattering response even at weak light-matter coupling $D\to0$. Panel (f) demonstrates this over a wider range of light-matter coupling strengths, for the linecut at $\wIC=0$, corresponding to the gray dotted lines in (c)--(e). As expected, the rotated response remains strong at exactly $\alpha=\pi/2$, even as $D\to0$. Note, however, that since the $g^{(2)}$ response is a ratio of two-photon vs one-photon detection, the total one and two photon counts individually will go to zero as $D\to0$. Fig. \ref{fig:polarization_rotation}(g) shows that the $\alpha=\pi/2$ rotated response also remains strong as we increase the linewidth $\gamma$.

\section{Outlook}\label{sec:outlook}

    In this paper, we show how the polarization-dependent photon statistics of light transmitted through a cavity with an embedded quantum magnet can be used for diagnosing material properties in cavity-embedded systems in transmission geometry. To this end, we study $g^{(2)}$ measurements at coincidence as a function of input and output photon polarization. First, polarization-resolved bunching and antibunching of output photons can resolve the magnetic point group symmetries of the material phase. To describe the observed input-frequency-dependent transitions between bunched and antibunched emission, we first study the output photon statistics at coincidence $g^{(2)}(0)$ for photons with the same polarization as the input field, and establish a relation to the Raman dynamical structure factor of the material. In particular, we show that input frequencies that are red-detuned from the lowest Raman-active matter excitation generically lead to antibunched emission, proportional to the Raman cross section. We find discontinuities in the coincidence response and polarization dependence across a first-order quantum phase transition that breaks rotation symmetries. Second, we investigate the $g^{(2)}$ coincidence response for output photon pairs whose polarization is \textit{orthogonal} to the input field. Here, polarization filtering at the detector selects photon pairs that both scattered off the material, providing a sensitive probe of cavity light-matter interactions. We show that the observed orthogonal $g^{(2)}$ can be related to higher-order matter correlation functions and characterize the dependence on light-matter coupling strength and cavity linewidths.

    One immediate extension of the presented polarization-resolved photon-statistics diagnostic is to study the Kitaev spin liquid (KSL) phase of the Kitaev-Heisenberg model. The KSL state could have a distinctive coincidence response coming from the continuum of fractionalized anyonic excitations that do not coherently contribute to two-photon resonant bunching. Furthermore, higher-order photon correlation functions \cite{nambiar2025diagnosing} as described in the polarization-rotated channel could potentially probe non-local and $\mathbb{Z}_2$ flux operators. We note that a faithful representation of the KSL excitation spectrum however requires access to larger system sizes inaccessible using exact diagonalization. In Kitaev-Heisenberg materials, including \ce{Na2IrO3}, \ce{\alpha-RuCl3}, and \ce{\alpha-Li2IrO3}, substantial effort has been focused on realizing the KSL ground state, including with cavities \cite{Chiocchetta2021,vinas2023controlling}. Although experimental evidence shows magnetic ordering at low temperatures \cite{Singh2010,Liu2011,Singh2012,Williams2016,Cao2016,Banerjee2017}, and applying strong magnetic fields or high pressure appears to access a spin-liquid-like regime \cite{Baek2017,Zheng2017,Stahl2024}, the precise nature of the observed phase remains ambiguous. Studying the polarization-dependent two-photon coincidence response could help diagnose whether a KSL state can be stabilized via embedding the material in a cavity.

    Beyond its use as a probe, the frequency and polarization-dependent antibunching has potential use for the generation of quantum light. In addition to the antibunched regions for cross-polarized output photons, intermediate polarization rotation angles can produce antibunching at the same frequencies where bunching occurs for orthogonal polarizations. Our decomposition of bunching and antibunching into Raman structure factors therefore informs how and in which materials such responses can be maximized. Doing so in quantum materials near a phase transition could provide a tunable polarization knob to realize light with non-classical photon statistics. Finally, we expect that our method should be a sensitive probe of polarization entanglement between photons measured by Bell-type inequalities. By tuning material phases and light-matter couplings, entangled photon pairs may be generated or transformed using cavities. The generation of antibunched single photons, the processing of quantum light and the measurement of photon statistics have potential applications in photon-based sensing, quantum information and computation \cite{flamini2019,gisin2007quantum,RevModPhys.89.035002,talkington2026b}.

\section{Acknowledgments}

    We thank D. P. Carmichael, S. Divic, and E. J. Mele for comments on the manuscript. B. K. and M. C. acknowledge support from Charles E. Kaufman Foundation under a New Initiative grant, the Alfred P. Sloan Foundation through a Sloan Research Fellowship, and the Center for Quantum Information, Engineering, Science and Technology (QUIEST) of the University of Pennsylvania. S. T. acknowledges support from the NSF under Grant No. DGE-1845298.

%


\appendix

\section{Kitaev Material Phases and Biasing}
\label{app:kitaevMaterialPhases}

    \begin{figure}
        \centering
        \includegraphics[width=1.0\columnwidth]{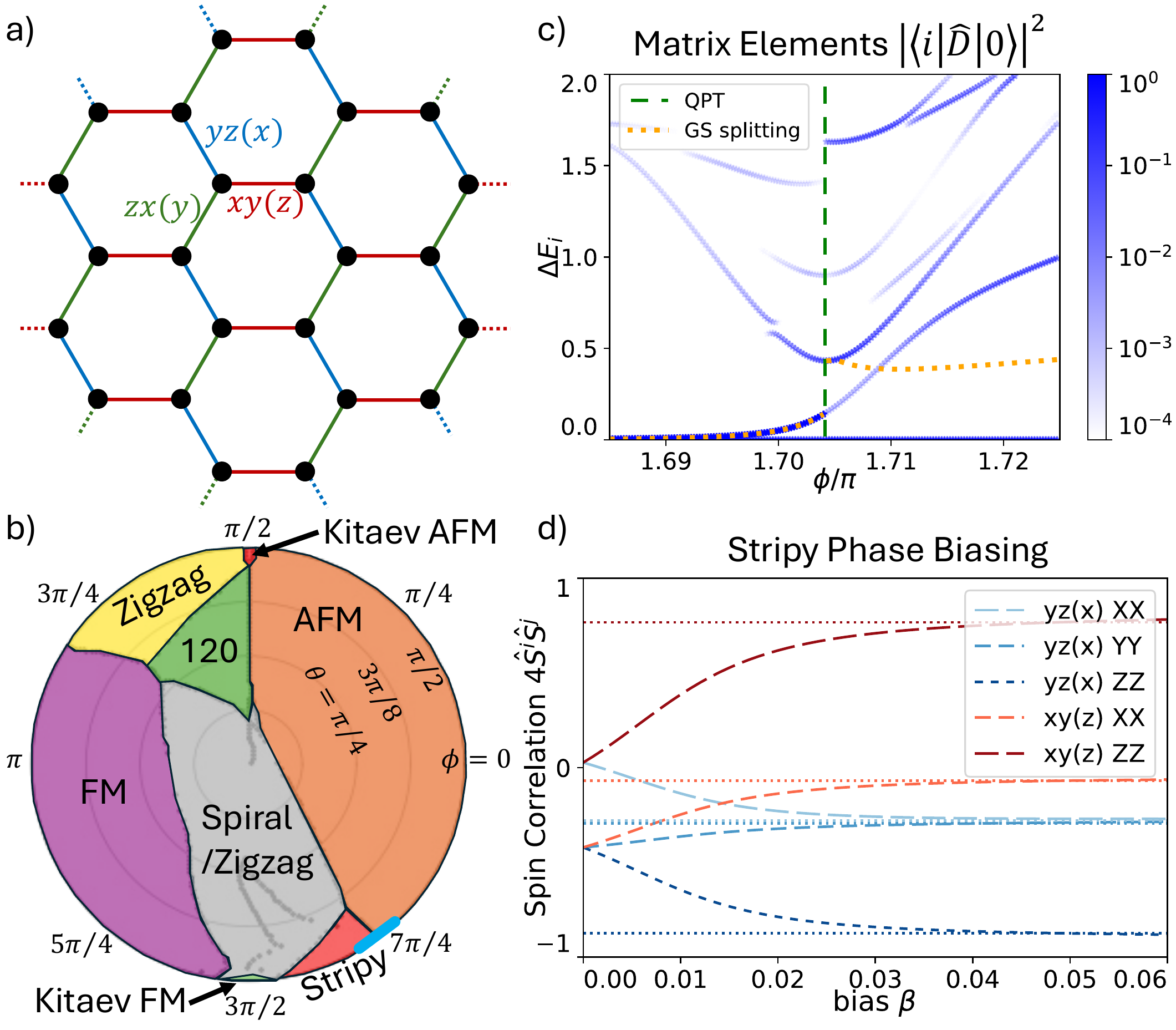}
        \caption{\textbf{Model and Biasing Info.} \textbf{(a)} $yz(x)$, $zx(y)$, and $xy(z)$ bonds of the $J$-$K$-$\Gamma$ model shown in blue, green, and red respectively on the 24 site cluster with periodic boundary conditions. \textbf{(b)} Phase diagram of the $J$-$K$-$\Gamma$ model with $\Gamma>0$, parameterized by $\theta$ and $\phi$ as in Eq. (\ref{eq:jkg_thetaphi}). The blue line shows the linecut used to study the stripy-to-antiferromagnetic (AFM) quantum phase transition (QPT). \textbf{(c)} Strength of the matrix elements from the ground state to Raman active excitations across the stripy-to-AFM QPT. The orange dotted lines show the finite size splitting of the ground state manifold; this lines up with the Raman active low energy excitations in the stripy phase, but sits at higher energy in the AFM phase. \textbf{(d)} Spin correlation $4\hat{S}^i\hat{S}^j$ for a selection of bond and spin directions as a biasing field $-\beta\hat{S}^z\hat{S}^z$ is turned on for the $xy(z)$ bonds at the point $(\theta,\phi)=(0.5,1.7)\pi$. At large biasing fields, all spin correlations are close to the equivalent values (dotted lines) for the ground state constructed as described in the text. }
        \label{fig:phase_info}
    \end{figure}

    For the purposes of fitting light-matter coupling, we used a full $J$-$K$-$\Gamma$ model, which extends the Kitaev-Heisenberg model with off-diagonal elements as
    \begin{align}
        \Ham &= \sum_{\langle ij\rangle} \begin{pmatrix}\hat{S}_i^\alpha & \hat{S}_i^\beta & \hat{S}_i^\gamma\end{pmatrix} \begin{pmatrix}J & \Gamma & \Gamma' \\ \Gamma & J & \Gamma' \\ \Gamma' & \Gamma' & J+K\end{pmatrix} \begin{pmatrix}\hat{S}_j^\alpha \\ \hat{S}_j^\beta \\ \hat{S}_j^\gamma\end{pmatrix} \notag \\
        &\qquad + B \sum_i \left( \hat{S}_i^x + \hat{S}_i^y + \hat{S}_i^z \right). \label{eq:khgHam}
    \end{align}
    Here, $\alpha$, $\beta$, and $\gamma$ are bond-dependent spin directions with $\alpha\beta(\gamma)\in\left\{yz(x),zx(y),xy(z)\right\}$ as specified in Fig. \ref{fig:phase_info}(a). $J$, $K$, and $\Gamma$/$\Gamma'$ are nearest-neighbor Heisenberg, Kitaev, and off-diagonal spin coupling terms. B is the applied magnetic field strength, which we choose to lie in the [111] direction in order to preserve the $S_6$ improper rotational symmetry. For fitting light-matter coupling, we use $\Gamma'=0$ and $\Gamma>0$, a region of phase space that has a rich and well-studied phase diagram \cite{rau2014}, as shown in Fig. \ref{fig:phase_info}(b). We follow the convention to normalize $J$, $K$, and $\Gamma$, such that $J^2+K^2+\Gamma^2=1$, and parametrize with $\theta$ and $\phi$ as
    \begin{align}
        J=\sin\theta\cos\phi,\quad K=\sin\theta\sin\phi,\quad \Gamma=\cos\theta. \label{eq:jkg_thetaphi}
    \end{align}
    Throughout this paper, we study the region around the stripy-to-AFM phase transition with $\theta=\pi/2$ and $1.685\pi\leq\phi\leq1.725\pi$, corresponding to the blue line in Fig. \ref{fig:phase_info}(b). Fig. \ref{fig:phase_info}(c) shows the expected Raman active excitations in this region, where the opacity of blue shows the strength of matrix elements $\left|\bra{i}\hat{D}\ket{0}\right|^2$ coupling each state $\ket{i}$ to the ground state $\ket{0}$. The strength of these matrix elements plays a critical role in the strength of the coincidence response.

    In Fig. \ref{fig:phase_info}(c), the orange dotted lines show the splitting of the ground state manifold caused by finite size effects on the 24 site cluster used. Instead of the magnetically ordered ground states expected in the thermodynamic limit, similar to the classical spin ordering shown in Fig. \ref{fig:model_and_phase}(a), the finite size ground state is a superposition of orientations. In the stripy phase, the superposition is between the three orientations in the zero momentum sector, which gives the ground state $C_3$ rotational symmetry. In the AFM phase, the superposition is between two states separated by a 180 degree rotation at the center of an $xy(z)$ bond, giving the ground state $C_6$ rotational symmetry. We choose to bias the stripy phase, since we expect qualitatively different polarization-dependent coincidence response between the $C_3$ superposition ground state and the $C_2$ biased state. We choose not to bias the AFM ground state, however, for two reasons. First, the finite size energy splitting is much larger in this phase, as seen by the orange dotted line in Fig. \ref{fig:phase_info}(c), which will have non-negligible impact on the excitation energies involved in the coincidence response. Second, we expect the $C_6$ superposition ground state to have qualitatively the same polarization-dependent coincidence response as the $C_3$ biased state, since linearly polarized photons behave as directors, which do not distinguish 180 degree rotations. A comparison of the coincidence response with and without biasing is given in Appendix \ref{app:biasingImpact}.

    One approach for breaking the symmetry of the ground state in finite size systems is to apply a directional bias to the material Hamiltonian. Fig. \ref{fig:phase_info}(d) shows the impact of applying a bias $-\beta\hat{S}^z\hat{S}^z$ along $xy(z)$ bonds for the stripy point $(\phi,\theta)=(1.7,0.5)\pi$. As expected, this causes the spins to align along the $xy(z)$ bonds into a stripy pattern, as confirmed by a selection of spin correlations shown as solid lines. A downside to this approach is that the bias size $\beta$ needs to be made large enough to overcome the finite size splitting of the ground state manifold, and doing so can introduce components of the biased ground state that did not originate in the original ground state manifold.

    Instead, we use an approach that does not depend on an arbitrary bias size. We construct the biased ground state as a linear combination of the three lowest energy eigenstates of $\HamMat$ in the zero momentum sector. We solve for the linear combination that maximizes $\hat{S}^z\hat{S}^z$ along $xy(z)$ bonds, while keeping $\hat{S}^z\hat{S}^z$ equal between the $yz(x)$ and $zx(y)$ bonds. The spin correlations of this state for the stripy point $(\phi,\theta)=(1.7,0.5)\pi$ are shown as dotted lines in Fig. \ref{fig:phase_info}(d). As expected, the spin correlations of the biased states converge closely to those of the constructed state. For large biases, the spin correlations actually overshoot slightly, which is caused by the contribution to the biased states from outside the ground state manifold, as mentioned above.

\section{Hopping Strengths for Light-Matter Coupling}
\label{app:lightMatterCouplingFitting}

    \begin{figure}
        \centering
        \includegraphics[width=1.0\columnwidth]{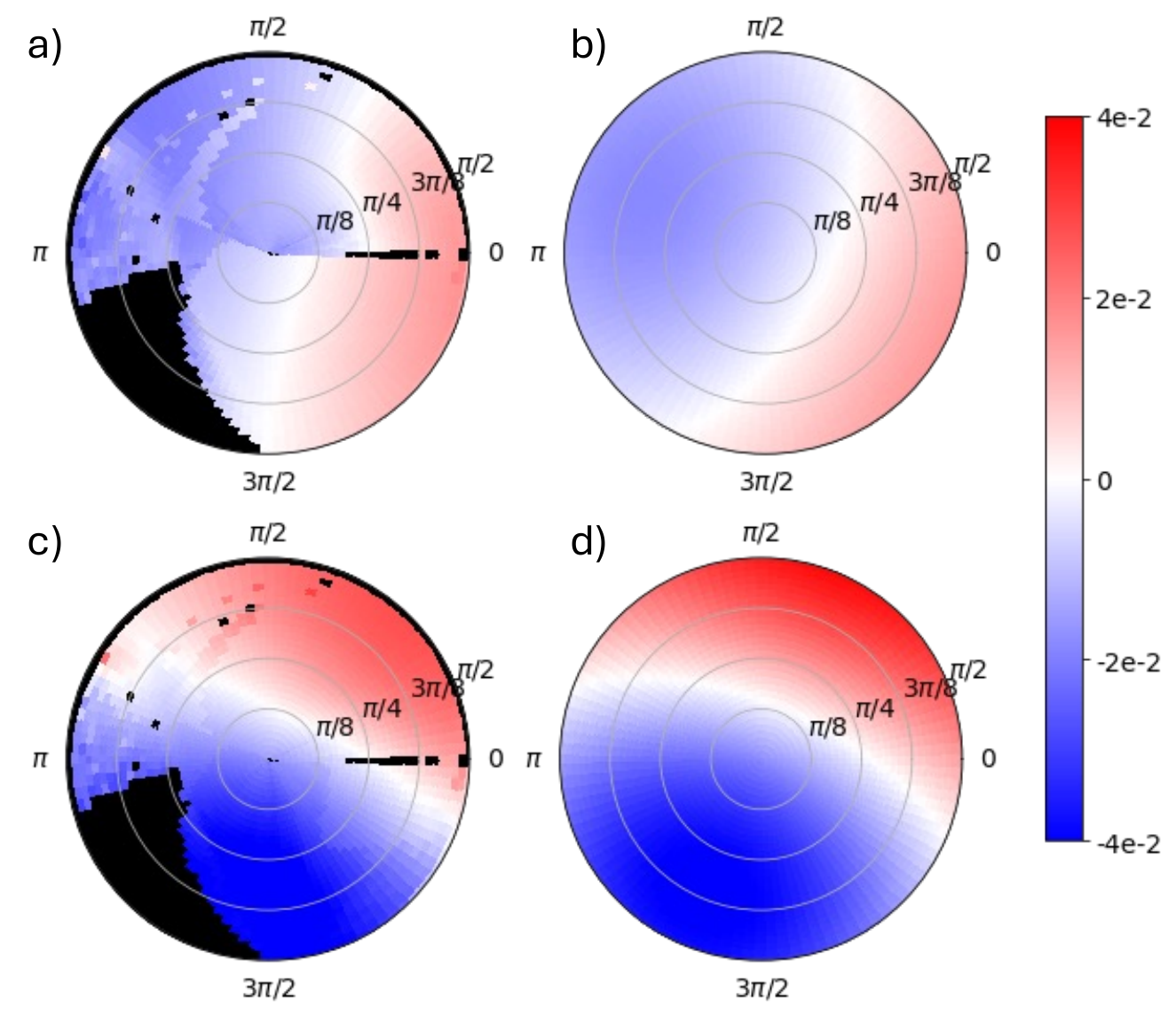}
        \caption{\textbf{Light-Matter Coupling Parameters.} $D^{11}_{xx}$ along the $xy(z)$ bonds \textbf{(a)} as calculated from quasi-degenerate perturbation theory going from the full to the spin Hamiltonian where possible (fitting errors above $10^{-6}$ appear in black) vs \textbf{(b)} using a least squares best fit against $J$, $K$, and $\Gamma$. \textbf{(c)} and \textbf{(d)} are the same for $D^{11}_{zz}$. }
        \label{fig:coupling_params}
    \end{figure}
    
    In order to estimate reasonable light-matter coupling $\Dop$ for Kitaev materials, we follow the approach used in \cite{vinas2023controlling} to estimate $J$, $K$, $\Gamma$, and $\Gamma'$ of the $J$-$K$-$\Gamma$ model of Eq. (\ref{eq:khgHam}) for \ce{\alpha-RuCl3}. This involves starting from a Hubbard-Kanamori Hamiltonian plus hopping terms for electrons in \ce{Ru^{3+}} d-orbitals, intermediated by \ce{Cl-} $p$-orbitals, corresponding to a single $xy(z)$ bond of the $J$-$K$-$\Gamma$ model. This Hamiltonian depends on intraorbital Hubbard interaction $U$, Hund's coupling $J_H$, spin-orbit coupling strength $\lambda$, $p$- vs $d$- orbital charge-transfer energy $\Delta_{pd}$, hopping strengths between different orbitals including $t_1$, $t_2$, $t_3$, $t_4$, and $t_{pd}$. We add two cavity modes parallel and perpendicular to the $xy(z)$ bonds, each with cavity mode frequency $\omega$. We then use a Peierls substitution to couple these modes to the hopping terms with a light-matter coupling strength $g$. The full Hamiltonian is down-folded to the spin sector using quasi-degenerate perturbation theory \cite{Winkler2003}. This can be done separately for each photon sector to get the contribution to $\HamMat$ and $\HamCavOne$ from the $xy(z)$ bonds. Inverting Eq. (\ref{eq:onePhotonSector}), we get
    \begin{align}
        \begin{pmatrix}
            \DMat{11} & \DMat{12} \\
            \DMat{21} & \DMat{22}
        \end{pmatrix} = \HamCavOne - \left(\HamMat+\omega_0\right)\otimes 1,
        \label{eq:onePhotonSectorInverted}
    \end{align}
    from which we calculate the light-matter coupling along the $xy(z)$ bonds, which take the form
    \begin{align}
        \DMat{ij}&=\sum_{\sigma\in\left\{x,y,z\right\}}D^{ij}_{\sigma\sigma}\hat{S}^\sigma\hat{S}^\sigma+D^{ij}_{xy}\left(\hat{S}^x\hat{S}^y+\hat{S}^y\hat{S}^x\right)
    \end{align}
    The new issue is that there is not a unique set of Hubbard-Kanamori parameters associated with each point $(\phi,\theta)$ of the $J$-$K$-$\Gamma$ model. To handle this, we fix the parameters $U=3.66$, $J_H=0.64$, $\lambda=0.18$, $\Delta_{pd}=4.34$, and $t_{pd}=-0.8$, all taken from the values calculated from first principles for \ce{\alpha-RuCl3} in \cite{vinas2023controlling}. We find $t_4$ is closely associated with $\Gamma'$, and since we focus on $\Gamma'=0$, we use $t_4=0$ instead of the value for \ce{\alpha-RuCl3}. Finally, we choose $g=0.3$ and $\omega=0.6$, with the constraints that $\omega$ should sit below the material charge gap to avoid photon absorption and $g$ should be large enough to have non-negligible coincidence response.

    After fixing these parameters, we are left with three remaining hopping terms $t_1$, $t_2$, and $t_3$, which we attempt to fit to $J$, $K$, and $\Gamma$ by minimizing the square error of the parameters of the down-folded spin model in the zero photon sector. As long as the total sum of squares error of this fit is less than a tolerance set to $10^{-6}$, we proceed to solve for $D^{ij}_{xx}$, $D^{ij}_{yy}$, $D^{ij}_{zz}$, and $D^{ij}_{xy}$ for the $xy(z)$ bonds. The Hubbard-Kanamori  parameters fixed to \ce{\alpha-RuCl3} are not flexible enough to fit all points in the $J$-$K$-$\Gamma$ model phase diagram (see the black regions in Figs. \ref{fig:coupling_params}(a) and (c)). Therefore, we perform a least squares fit for $D^{ij}_{xx}$, $D^{ij}_{yy}$, $D^{ij}_{zz}$, and $D^{ij}_{xy}$ against $J$, $K$, and $\Gamma$ on a grid of points. These best fit coefficients are used to predict $D^{ij}_{xx}$, $D^{ij}_{yy}$, $D^{ij}_{zz}$, and $D^{ij}_{xy}$ across the entire phase diagram. A comparison of the direct calculation vs least squares fit for $D^{11}_{xx}$ and $D^{11}_{zz}$ along $xy(z)$ bonds is shown in Fig. \ref{fig:coupling_params}. Finally, these parameters are rotated in polarization and spin to get the equivalent coupling terms for the $yz(x)$ and $zx(y)$ bonds.

\section{Cavity Mode Frequency}
\label{app:cavityModeFrequency}

    \begin{figure}
        \centering
        \includegraphics[width=1.0\columnwidth]{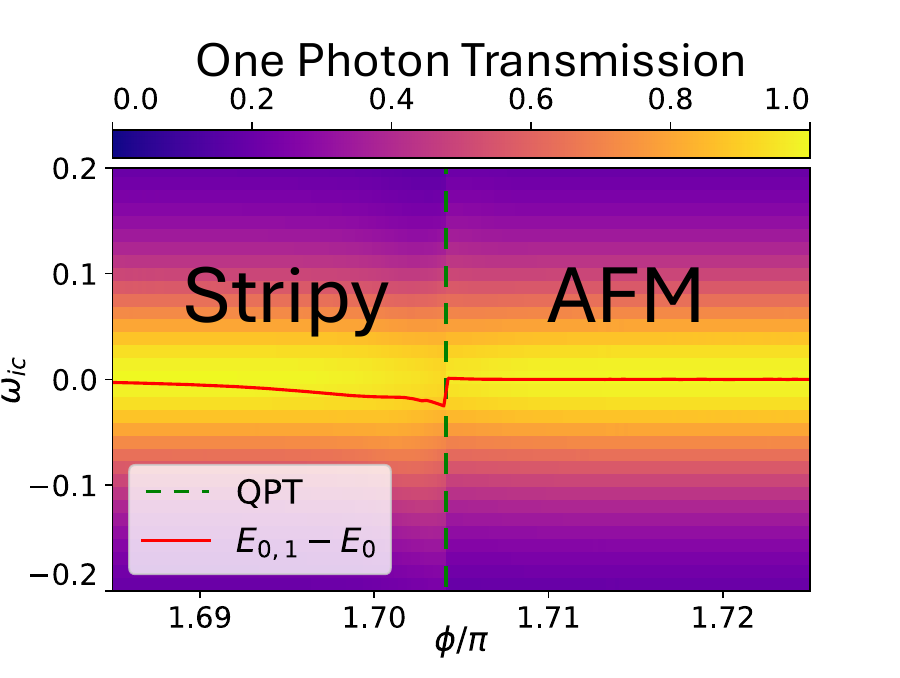}
        \caption{\textbf{One Photon Transmission.} $G^{(1)}(\omega_{\rm{ic}})$ for a single polarization across the stripy-to-antiferromagnetic (AFM) phase transition with the same parameters as in Fig. \ref{fig:model_and_phase}(d). The maximum transmission occurs at $\omega_{\rm{ic}}=0$. The red line shows $\omega_c'=E_{0,1}-E_0$, at which the input frequency $\wIn=\omega_c'$ is one-photon resonant with $\ket{gs,1}$. The green dashed line shows the quantum phase transition (QPT). }
        \label{fig:photon_transmission}
    \end{figure}
    
    The choice of how to define the cavity mode frequency $\omega_c$ does not change the coincidence spectrum, but can help by centering the most interesting features of the spectrum near $\omega_{\rm{ic}}=0$. We find that a good choice is to define $\omega_c=E_{0,1}-E_0$ where $E_0$ is the ground state energy of $\HamMat$ and $E_{0,1}$ is the ground state energy of $\HamCavOne$. With this convention, $\omega_{\rm{ic}}=0$ will line up with the one photon resonance with $\ket{gs,1}$, since the denominator of Eq. (\ref{eq:G1Definition}) becomes
    \begin{align}
        \wIn+E_0-\HamCavOne+i\gamma&\to\omega_{\rm{c}}+E_0-E_{0,1}+i\gamma\notag\\
        &\to i\gamma,
    \end{align}
    where in the first line we set $\omega_{\rm{in}}\to\omega_c$ and $\HamCavOne$ to its ground state energy, and in the second line we used the above convention for $\omega_c$.

    In practice, however, we choose to approximate this result with experimentally available data by choosing $\omega_c$ based on the frequency that maximizes one photon transmission $G^{(1)}(\omega_{\rm{ic}})$, which corresponds to each part of the denominator in Eq. (\ref{eq:g2withpolarization}) up to factors of $\gamma$. In the single-polarization case
    \begin{align}
        G^{(1)}(\omega_{\rm{ic}}) = \gamma^2\sum\limits_{f} \left| \bra{f} \GTen_1 \polVec_{\rm{in}}\ket{0} \right|^2. \label{eq:onePhotonTransmissionNoPolarization}
    \end{align}
    Fig. \ref{fig:photon_transmission} shows $G^{(1)}(\omega_{\rm{ic}})$ across the stripy-to-AFM QPT with the same parameters as in Fig. \ref{fig:model_and_phase}(d). By construction, we see that the maximum value of $G^{(1)}(\omega_{\rm{ic}})$ is at $\omega_{\rm{ic}}=0$. The red line shows that the value $\omega_c'=E_{0,1}-E_0$ is very similar. The largest difference occurs in the stripy phase, due to our construction of the broken symmetry stripy state, which is not exactly the ground state in our finite sized system \seeAppendixX{\ref{app:kitaevMaterialPhases}}. Smaller differences come from the overlap of nearby resonances.

    With polarization, we perform the same procedure to choose $\omega_c$ to maximize $G_\mu^{(1)}(\omega_{\rm{ic}},\nu)$. Now, however, the value can depend on the input and output polarizations. We choose the arbitrary convention to define $\omega_c$ relative to linearly polarized input and output photons that are aligned with the Kitaev $z$ bonds.

\section{Coincidence Perturbative Expansion}
\label{app:perturbativeExpansion}

    \begin{figure}
        \centering
        \includegraphics[width=\linewidth]{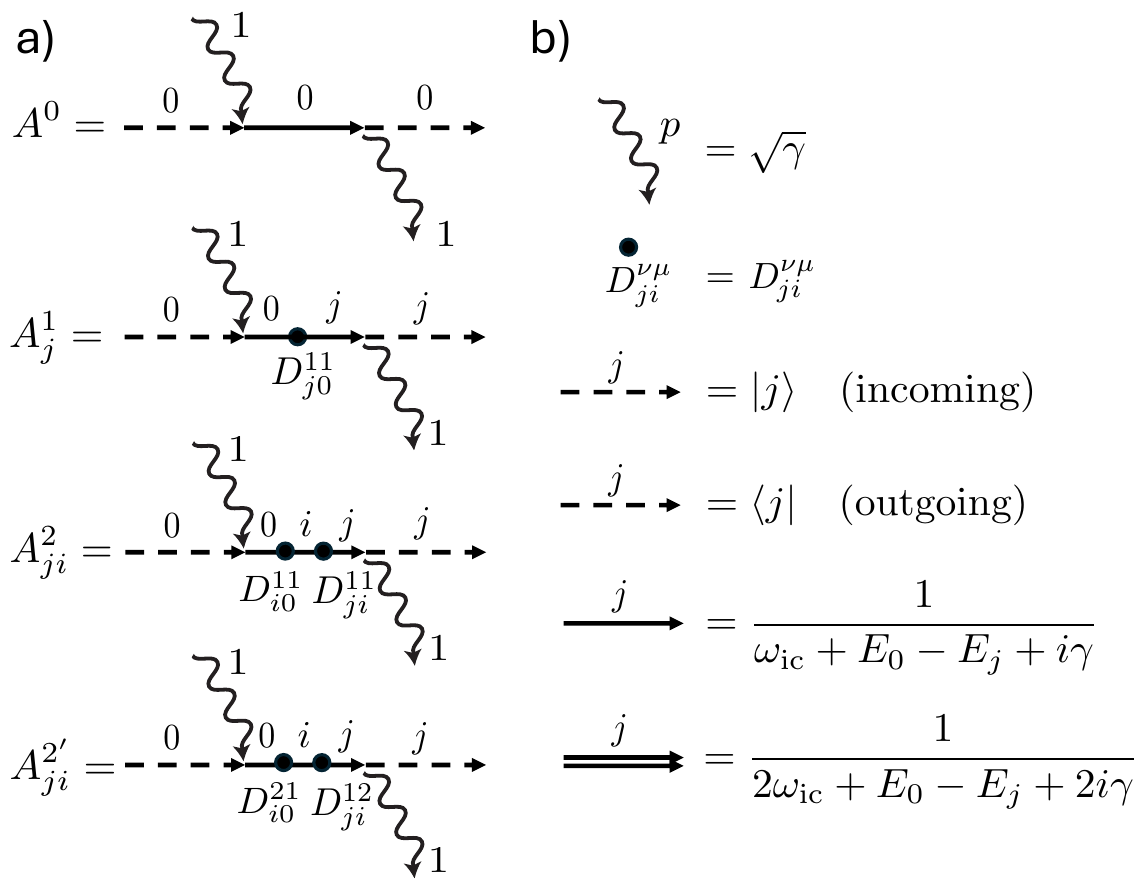}
        \caption{(a) Diagrams for amplitudes contributing to the unrotated $G^{(1)}$ response at order $D^2$. (b) Rules for the scattering processes we consider. The squiggly line is an incoming/outgoing photon with polarization $p$. The dot is a cavity photon-matter scattering. The dashed line with $j$ above is the cavity, with material state $j$, which can either be a ket (incoming from the left) or a bra (outgoing to the right). The solid line with $j$ is the cavity with one photon and material excited into state $j$. The double solid line is the cavity with two photons and material excited into state $j$.}
        \label{fig:diagram-G1}
    \end{figure}

    \begin{figure*}
        \centering
        \includegraphics[width=\linewidth]{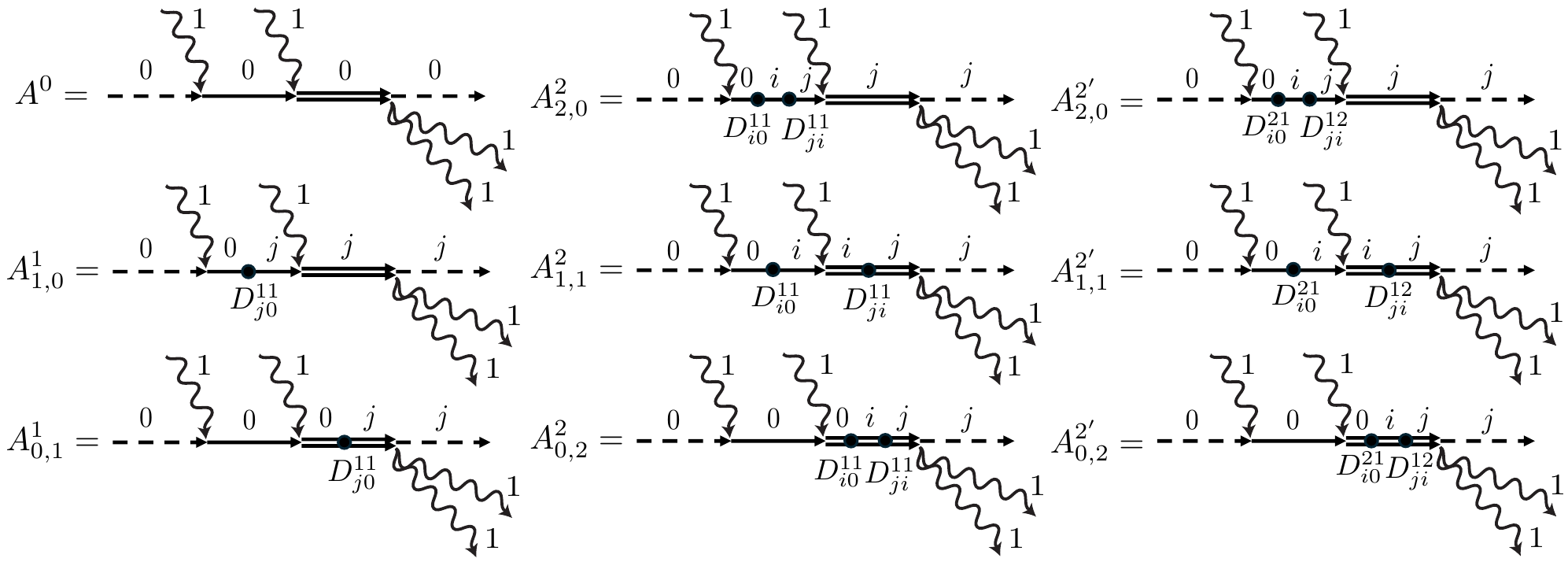}
        \caption{Scattering amplitudes $A$ expressed as diagrams for two-photon scattering off a cavity embedded material corresponding to unrotated light as expressed in Eq. (\ref{eq:G2-mu=1-D2}). Notably there is no need for a second single photon resolvent (as its contribution is just the identity in the Markov approximation \cite{kass2024}) so both output photons can be drawn as being simultaneously emitted. We take $j$ and $ji$ indices to be implicitly present on $A$ where $A^1$ carries a $j$ index and $A^2$ carries $ji$ indices. Diagram constituents are as in Fig. \ref{fig:diagram-G1}.}
        \label{fig:diagram-G2}
    \end{figure*}

    We first derive a perturbative expansion for the polarization-dependent one photon transmission
    \begin{align}
        G_\mu^{(1)}(\omega_{\rm{in}},\nu) = \gamma^2\sum\limits_{f} \left| \bra{f} \polVecC_\mu \GTen_1 \polVec_{\rm{in},\nu} \otimes \ket{0} \right|^2, \label{eq:onePhotonTransmission}
    \end{align}
    which is one part of the denominator in Eq. (\ref{eq:g2withpolarization}) up to powers of $\gamma$. For convenience, we use a polarization basis 1 and 2 where $1=\nu$, we define $\omega_c=\omega_0+\bra{0}\DMat{11}\ket{0}$ and redefine $\DMat{11}\to\DMat{11}-\bra{0}\DMat{11}\ket{0}$. Then, for small $D$, we can write the Green's resolvent $\GTen_1$ as a Neumann series expansion
    \begin{align}
        \GTen_1 &= \sum_{n=0}^\infty \left(\GTen_1^0\Dop_1\right)^n \GTen_1^0
    \end{align}
    where
    \begin{align}
        \GTen_1^0 &= \Gop_1^0\otimes1\\
        \Gop_1^0&=\frac{1}{\wIC +E_0 - \HamMat + i\gamma} \\
        \Dop_1&=\begin{pmatrix}
            \DMat{11} & \DMat{12} \\
            \DMat{21} & \DMat{22}
        \end{pmatrix}.
    \end{align}
    First, for unrotated transmission, we get a $D^0$ term $\delta_{\mu,\nu}\gamma^2\bra{0}\Gop_1^{0\dagger}\Gop_1^0\ket{0}=\delta_{\mu,\nu}\gamma^2/\left(\wIC^2+\gamma^2\right)$. Since $\bra{0}\DMat{11}\ket{0}=0$, we do not have any terms of order $D$, so we next list out all possible terms with two factors of $\DMat{}$. The rotated response is
    \begin{align}
        &\hspace{-0.48 in} \includegraphics[width=0.62\linewidth]{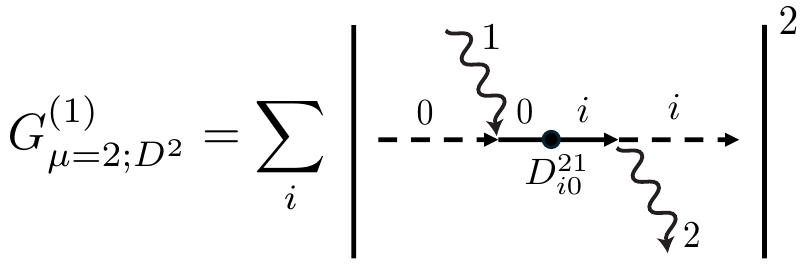}\\
        &=\gamma^2\bra{0}\Gop_1^{0\dagger}\DMat{12}\Gop_1^{0\dagger}\Gop_1^0\DMat{21}\Gop_1^0\ket{0}
    \end{align}
    and the unrotated response is, in terms of the amplitudes in Fig. \ref{fig:diagram-G1}(a)
    \begin{align}
        G_{\mu=1;D^2}^{(1)} &= \sum\nolimits_j \big[A^{1\dag}_jA^1_j + A^{0\dag}(\sum\nolimits_i A^2_{ji}+A^{2'}_{ji})\\&\hspace{1.14 in} + (\sum\nolimits_i A^2_{ji}+A^{2'}_{ji})^\dag A^0\big]\notag\\
        &=\gamma^2\bra{0}\bigg(\Gop_1^{0\dagger}\DMat{11}\Gop_1^{0\dagger}\Gop_1^0\DMat{11}\Gop_1^0\\
        &+\Gop_1^{0\dagger}\Gop_1^0\DMat{11}\Gop_1^0\DMat{11}\Gop_1^0+\Gop_1^{0\dagger}\Gop_1^0\DMat{12}\Gop_1^0\DMat{21}\Gop_1^0\notag\\
        &+\Gop_1^{0\dagger}\DMat{11}\Gop_1^{0\dagger}\DMat{11}\Gop_1^{0\dagger}\Gop_1^0+\Gop_1^{0\dagger}\DMat{12}\Gop_1^{0\dagger}\DMat{21}\Gop_1^{0\dagger}\Gop_1^0\bigg)\!\ket{0} \notag
    \end{align}
    Inserting a complete basis in between the two factors of $\hat{D}$ and simplifying in terms of excitation energies $\Delta E_i=E_i-E_0$ and matrix elements $D^{ij}_{i0}=\bra{i}\DMat{ij}\ket{0}$ for eigenstates $\ket{i}$ of $\HamMat$, we get that the second order perturbative expansion for the unrotated one photon transmission is
    \begin{align}
        G_{\mu=1}^{(1)} &= \frac{\gamma^2}{\wIC^2+\gamma^2}\bigg(1\,+\sum_i\frac{\left(3\wIC^2-2\wIC\Delta E_i-\gamma^2\right)D^{11}_{0i}D^{11}_{i0}}{\left(\wIC^2+\gamma^2\right)\left(\left(\wIC-\Delta E_i\right)^2+\gamma^2\right)}\notag\\
        &+\left.\sum_i\frac{2\left(\wIC^2-\wIC\Delta E_i-\gamma^2\right)D^{12}_{0i}D^{21}_{i0}}{\left(\wIC^2+\gamma^2\right)\left(\left(\wIC-\Delta E_i\right)^2+\gamma^2\right)}\right)+\mathcal{O}(D^3)
        \label{app:unrotatedOnePhotonTransmissionPerturbative}
    \end{align}
    while the second order perturbative expansion for the rotated one photon transmission is
    \begin{align}
        G_{\mu=2}^{(1)} &= \!\frac{\gamma^2}{\wIC^2+\gamma^2}\sum_i\frac{D^{12}_{0i}D^{21}_{i0}}{\left(\wIC\!-\!\Delta E_i\right)^2\!+\gamma^2}+\mathcal{O}(D^3).
        \label{app:rotatedOnePhotonTransmissionPerturbative}
    \end{align}
    Next, we repeat this process for the polarization-dependent two photon transmission
    \begin{align}
        G_\mu^{(2)}(\omega_{\rm{in}},\nu) = \gamma^4\sum\limits_{f} \left| \bra{f} \polTensorC_{\mu\mu} \GTen_2 \polVec_{\rm{in},\nu} \otimes \left[ \GTen_1 \polVec_{\rm{in},\nu} \otimes \ket{0} \right] \right|^2, \label{eq:twoPhotonTransmission}
    \end{align}
    which is the numerator in Eq. (\ref{eq:g2withpolarization}) up to powers of $\gamma$. Again we use a Neumann series expansion to write
    \begin{align}
        \GTen_2 &= \sum_{n=0}^\infty \left(\GTen_2^0\Dop_2\right)^n \GTen_2^0
    \end{align}
    where
    \begin{align}
        \GTen_2^0 &= \Gop_2^0\otimes1\otimes1\\
        \Gop_2^0&=\frac{1}{2\wIC + E_0 - \HamMat + 2i\gamma}\\
        \Dop_2&=\begin{pmatrix}
            2\DMat{11} & \DMat{12} & \DMat{12} & 0 \\
            \DMat{21} & \DMat{11}+\DMat{22} & 0 & \DMat{12} \\
            \DMat{21} & 0 & \DMat{11}+\DMat{22} & \DMat{12} \\
            0 & \DMat{21} & \DMat{21} & 2\DMat{22}
        \end{pmatrix}.
    \end{align}
    
    The $D^0$ unrotated two photon transmission term is $4\delta_{\mu,\nu}\gamma^4\bra{0}\Gop_1^{0\dagger}\Gop_2^{0\dagger}\Gop_2^0\Gop_1^0\ket{0}=\delta_{\mu,\nu}\gamma^4/\left(\wIC^2+\gamma^2\right)^2$, where the factor of 4 came from $\polTensorC_{\mu\mu}$.
    There are 16 terms that contribute to the unrotated two photon transmission to order $D^2$; these can be understood as the interference terms of the diagrammatic amplitudes illustrated in Fig. \ref{fig:diagram-G2}---4 terms involve the interference of amplitudes $A^1$ with one material interaction with another $A^1$, 6 terms involve interference of the no-material interaction amplitude $A^0$ with a two-material interaction amplitude $A^2$ not involving an intermediate rotated state, and 6 terms involve interference of the no-material interaction amplitude $A^0$ with a two-material interaction amplitude $A^{2'}$ with an intermediate rotated state. Explicitly
\begin{widetext}
    \begin{align}\label{eq:G2-mu=1-D2}
        G_{\mu=1;D^2}^{(2)}
        = \sum\nolimits_j \big[(A&^1_{1,0}+A^1_{0,1})^\dag (A^1_{1,0}+A^1_{0,1}) +  (\sum\nolimits_i A^2_{2,0} + A^2_{1,1} + A^2_{0,2})^\dag A^0 + A^{0\dag} (\sum\nolimits_i A^2_{2,0} + A^2_{1,1} + A^2_{0,2}) \notag\\
        &\hspace{1.48 in} +  (\sum\nolimits_i A^{2'}_{2,0} + A^{2'}_{1,1} + A^{2'}_{0,2})^\dag A^0 + A^{0\dag} (\sum\nolimits_i A^{2'}_{2,0} + A^{2'}_{1,1} + A^{2'}_{0,2})\big]\\
        =\gamma^4\bra{0}\bigg(&(\Gop_1^{0\dagger}\DMat{11}\Gop_1^{0\dagger}\Gop_2^{0\dagger} + \Gop_1^{0\dag}\Gop_2^{0\dag}\DMat{11}\Gop_2^{0\dag})(\Gop_2^0\Gop_1^0\DMat{11}\Gop_1^0 + \Gop_2^0\DMat{11}\Gop_2^0\Gop_1^0) \notag\\
        &+(\Gop_1^{0\dagger}\DMat{11}\Gop_1^{0\dagger}\DMat{11}\Gop_1^{0\dagger}\Gop_2^{0\dagger} + \Gop_1^{0\dagger}\DMat{11}\Gop_1^{0\dagger}\Gop_2^{0\dagger}\DMat{11}\Gop_2^{0\dagger} + \Gop_1^{0\dagger}\Gop_2^{0\dagger}\DMat{11}\Gop_2^{0\dagger}\DMat{11}\Gop_2^{0\dagger})\ \Gop_2^0\Gop_1^0\notag\\
        &+\Gop_1^{0\dagger}\Gop_2^{0\dagger}\, (\Gop_2^0\Gop_1^0\DMat{11}\Gop_1^0\DMat{11}\Gop_1^0 + \Gop_2^0\DMat{11}\Gop_2^0\Gop_1^0\DMat{11}\Gop_1^0 + \Gop_2^0\DMat{11}\Gop_2^0\DMat{11} \Gop_2^0\Gop_1^0) \notag\\
        &+(\Gop_1^{0\dagger}\DMat{12}\Gop_1^{0\dagger}\DMat{21}\Gop_1^{0\dagger}\Gop_2^{0\dagger} + \Gop_1^{0\dagger}\DMat{12}\Gop_1^{0\dagger}\Gop_2^{0\dagger}\DMat{21}\Gop_2^{0\dagger} + \Gop_1^{0\dagger}\Gop_2^{0\dagger}\DMat{12}\Gop_2^{0\dagger}\DMat{21}\Gop_2^{0\dagger})\ \Gop_2^0\Gop_1^0\notag\\
        &+\Gop_1^{0\dagger}\Gop_2^{0\dagger}\,(\Gop_2^0\Gop_1^0\DMat{12}\Gop_1^0\DMat{21}\Gop_1^0 + \Gop_2^0\DMat{12}\Gop_2^0\Gop_1^0\DMat{21}\Gop_1^0 +\Gop_2^0\DMat{12}\Gop_2^0\DMat{21}\Gop_2^0\Gop_1^0)\bigg)\ket{0},
    \end{align}
    but to rotate both photons, we need four powers of $D$. There are only four such terms, because at least one interaction must occur after the second photon enters the cavity
    \begin{align}
        &\hspace{-1.15 in}\includegraphics[width=0.7\linewidth]{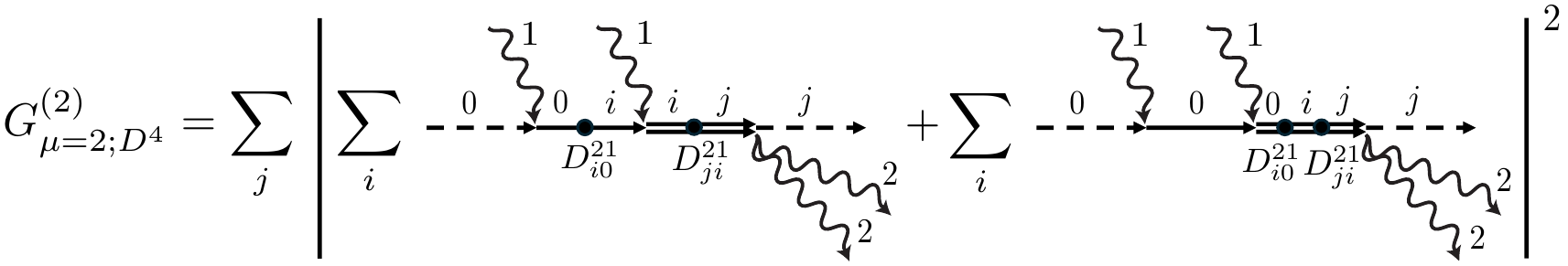}\\
        \phantom{ }\hspace{0.4 in} =\gamma^4\bra{0}\bigg(&\Gop_1^{0\dagger}\DMat{12}\Gop_1^{0\dagger}\Gop_2^{0\dagger}\DMat{12}\Gop_2^{0\dagger}\Gop_2^0\DMat{21}\Gop_2^0\Gop_1^0\DMat{21}\Gop_1^0+\Gop_1^{0\dagger}\DMat{12}\Gop_1^{0\dagger}\Gop_2^{0\dagger}\DMat{12}\Gop_2^{0\dagger}\Gop_2^0\DMat{21}\Gop_2^0\DMat{21}\Gop_2^0\Gop_1^0\notag\\
        &+\Gop_1^{0\dagger}\Gop_2^{0\dagger}\DMat{12}\Gop_2^{0\dagger}\DMat{12}\Gop_2^{0\dagger}\Gop_2^0\DMat{21}\Gop_2^0\Gop_1^0\DMat{21}\Gop_1^0+\Gop_1^{0\dagger}\Gop_2^{0\dagger}\DMat{12}\Gop_2^{0\dagger}\DMat{12}\Gop_2^{0\dagger}\Gop_2^0\DMat{21}\Gop_2^0\DMat{21}\Gop_2^0\Gop_1^0\bigg)\ket{0}.
    \end{align}
    We can insert a complete basis between every instance of $\DMat{}$ and simplify as before to get the perturbative expansion for the unrotated two photon transmission
    \begin{align}
        G_{\mu=1}^{(2)} &= \frac{\gamma^4}{\left(\wIC^2+\gamma^2\right)^2}\left(1\,+\sum_i\frac{\left(10\wIC^2-6\wIC\Delta E_i-2\gamma^2\right)D^{11}_{0i}D^{11}_{i0}}{\left(\wIC^2+\gamma^2\right)\left(\left(\wIC-\Delta E_i\right)^2+\gamma^2\right)}+\sum_i\frac{4\left(\wIC^2-\wIC\Delta E_i-\gamma^2\right)D^{12}_{0i}D^{21}_{i0}}{\left(\wIC^2+\gamma^2\right)\left(\left(\wIC-\Delta E_i\right)^2+\gamma^2\right)}\right)+ \mathcal{O}(D^3)
        \label{app:unrotatedTwoPhotonTransmissionPerturbative}
    \end{align}
    and the rotated two photon transmission
    \begin{align}
        G_{\mu=2}^{(2)} &= \frac{\gamma^4}{\left(\wIC^2+\gamma^2\right)^2} \sum_{jkl}\frac{D^{12}_{0j}D^{12}_{jk}D^{21}_{kl}D^{21}_{l0}}{\left(\wIC-\Delta E_j-i\gamma\right)\left(\left(\wIC-\Delta E_k/2\right)^2+\gamma^2\right)\left(\wIC-\Delta E_l+i\gamma\right)}+\mathcal{O}(D^5)
        \label{app:rotatedTwoPhotonTransmissionPerturbative}
    \end{align}
    \end{widetext}
    We then take the ratios of the above to get $g^{(2)}=G^{(2)}/(G^{(1)})^2$. Specifically, Eqs. (\ref{eq:singlePolarizationCoincidenceExpansion}), (\ref{eq:singlePolarizationCoincidenceExpansionCorrelators}) and (\ref{eq:unrotatedCoincidencePerturbativeCorrelation}) come from the ratio of Eqs. (\ref{app:unrotatedTwoPhotonTransmissionPerturbative}) / (\ref{app:unrotatedOnePhotonTransmissionPerturbative})$^2$ and Eqs. (\ref{eq:rotatedCoincidencePerturbative}) and (\ref{eq:rotatedCoincidencePerturbativeCorrelation}) come from Eqs. (\ref{app:rotatedTwoPhotonTransmissionPerturbative}) / (\ref{app:rotatedOnePhotonTransmissionPerturbative})$^2$.

\section{Single- vs Two-Polarization Mode Comparison}
\label{app:comparisonOfOneAndTwo}

    \begin{figure}[h]
        \centering
        \includegraphics[width=1.0\columnwidth]{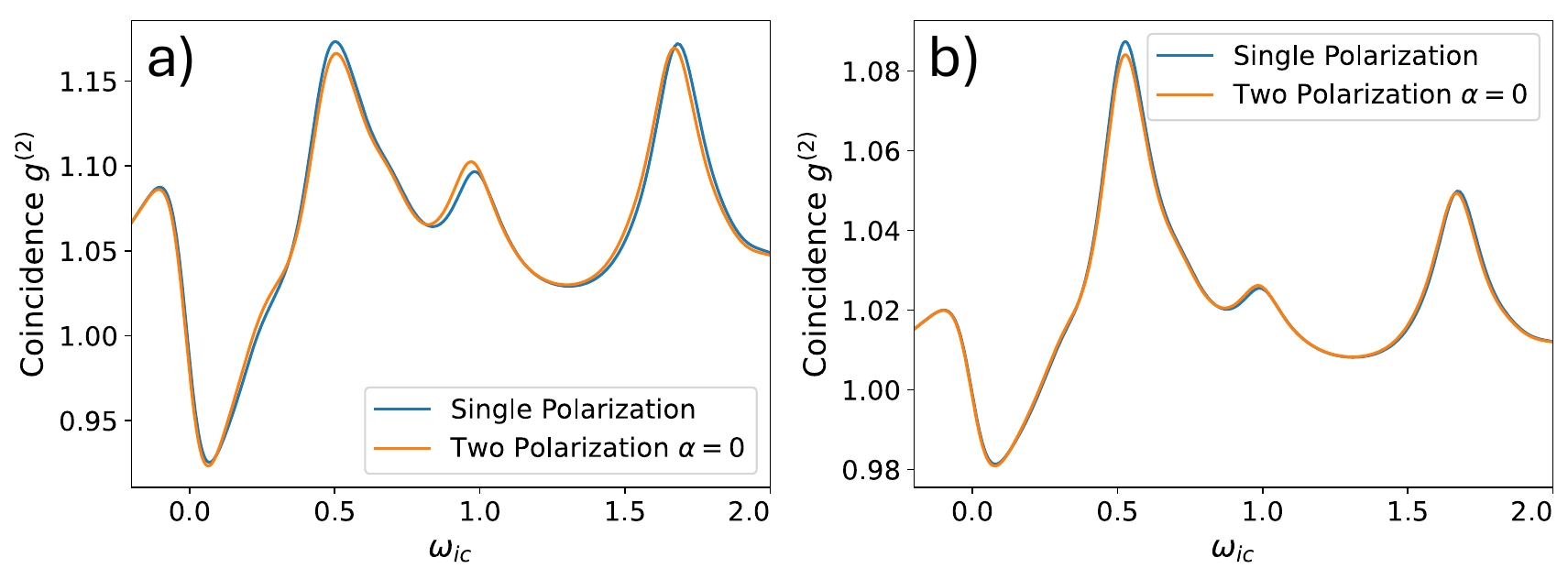}
        \caption{\textbf{Single vs Two Polarizations.} Comparison of single-polarization to two-polarization coincidence measurements when the input and output photons are all linearly polarized, aligned with the $xy(z)$ bonds, for $D$ scaled by \textbf{(a)} 1.0 and \textbf{(b)} 0.5. The coincidence spectrum is very similar, with the differences scaling as $D^4$. }
        \label{fig:single_versus_two}
    \end{figure}
    
    Comparing Eqs. (\ref{eq:singlePolarizationCoincidenceExpansionCorrelators}) and (\ref{eq:unrotatedCoincidencePerturbativeCorrelation}), we expect the single- and two-polarization coincidence measurements to be the same up to order $D^2$. Any differences enter at order $D^4$ and come from a photon in the two-polarization case rotating to the other polarization and back. Therefore, the corrections are small until $D$ reaches non-perturbative values. Fig. \ref{fig:single_versus_two}(a) confirms this intuition, showing small differences, while (b) further shows that the differences become even smaller when the light-matter coupling strength is halved.

\section{Comparison of Coincidence With and Without Ground State Biasing}
\label{app:biasingImpact}

    \begin{figure}
        \centering
        \includegraphics[width=1.0\columnwidth]{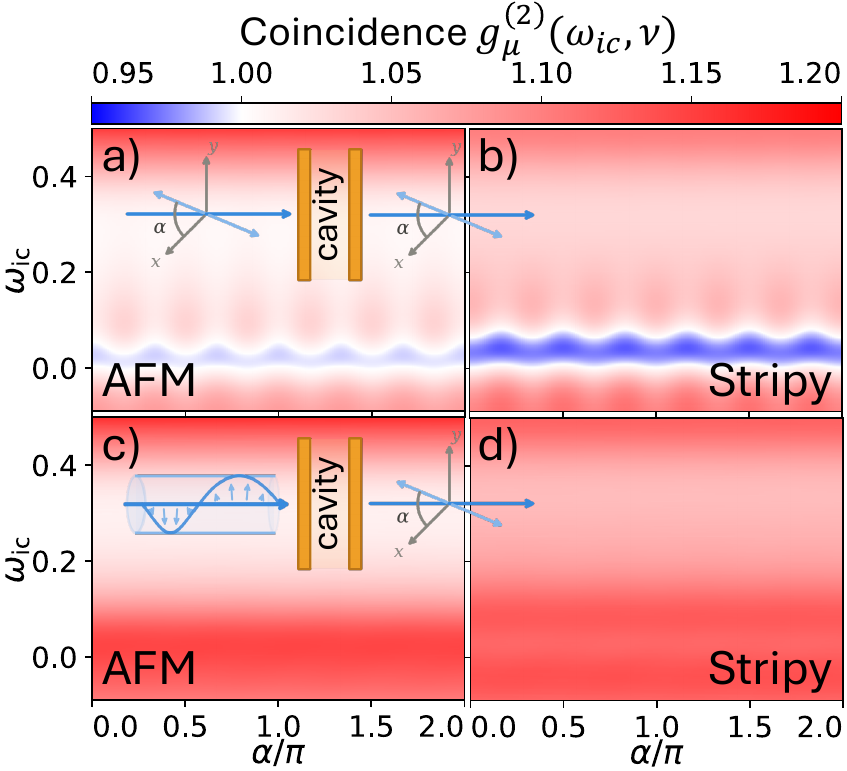}
        \caption{\textbf{Polarization-Dependent Coincidence With Opposite Biasing.} The same coincidence response to linearly/circularly polarized input photons as in Fig. \ref{fig:two_polarization_results}, but with different choices for biasing. \textbf{(a)} and \textbf{(c)} Coincidence for the AFM phase at $\phi=1.708\pi$ after constructing a biased initial state with $C_3$ rotational symmetry. \textbf{(b)} and \textbf{(d)} Coincidence for the stripy phase at $\phi=1.7\pi$ from the superposition ground state with $C_3$ rotational symmetry. The top row [(a),(b)] shows $C_6$ coincidence response to linearly polarized input photons that rotate at the same angle $\alpha$ as the output filter. The bottom row [(c),(d)] shows continuous coincidence response to right-circularly polarized input.}
        \label{fig:two_polarization_results_unbiased}
    \end{figure}

    Fig. \ref{fig:two_polarization_results_unbiased} shows the same coincidence response as in Fig. \ref{fig:two_polarization_results}, except flipping the choice of which state should be biased. Here, we look at the biased AFM state and the unbiased stripy state. Both of these have $C_3$ rotational symmetry. As expected, however, the linearly polarized input photons of panels (a) and (b) cannot distinguish $C_3$ from $C_6$, and the polarization-dependent coincidence response shows $C_6$ rotational symmetry. In both cases, the response to circularly polarized input photons has continuous rotational symmetry \seeAppendixX{\ref{app:continuousRotationalSymmetry}}. These results confirm that biasing is required for the stripy phase, but not for the AFM phase, in order to match the expected symmetries for polarization-dependent coincidence response in the thermodynamic limit. Comparing panels (a) and (c) between Figs. \ref{fig:two_polarization_results} and \ref{fig:two_polarization_results_unbiased}, we also see that the region of antibunching is much smaller after constructing a biased AFM ground state. This comes from the large finite size ground state energy splitting in this phase, shown by the orange line in Fig. \ref{fig:phase_info}(c); the coincidence response picks up resonances with much smaller excitation energies relative to the orange line. We expect that constructing a biased stripy state has a similar effect, but the energy splitting is much smaller in the stripy phase and is only visibly discernible at points near the QPT in Fig. \ref{fig:phase_info}(c).

\section{Continuous Rotationally Symmetric Response From Circularly Polarized Input}
\label{app:continuousRotationalSymmetry}

    The photon-dependent coincidence spectrum for circularly polarized input light will have continuous rotational symmetry for any initial state with $C_3$ rotational symmetry, as in Fig. \ref{fig:two_polarization_results}(c). To understand this, consider the two photon transmission rate as a response tensor in terms of input photon polarizations $\nu$ and $\nu'$, and output photon polarizations $\mu$ and $\mu'$. For each photon, the response tensor has a pair of indices of length 2, which can rotate into each other. Therefore, the response tensor $\tensor{T}{^i_j^k_l^m_n^o_p}$ is a $(2\times2)\times(2\times2)\times(2\times2)\times(2\times2)$ tensor with $(i,j)$ corresponding to $\nu$, $(k,l)$ corresponding to $\nu'$, $(m,n)$ corresponding to $\mu$, and $(o,p)$ corresponding to $\mu'$. If we express $T$ in the basis of right- ($R$) and left- ($L$) circular polarization, then rotating the system by an angle $\Theta$ will just add a phase to each tensor element. Upper indices contribute $e^{-i\Theta}$ for $R$ and $e^{+i\Theta}$ for $L$, while lower indices are reversed. In total, tensor elements can pick up phase $e^{xi\Theta}$ for $x\in\{0,\pm2,\pm4,\pm6,\pm8\}$. For example, the element $\phantom{\bigg(}\tensor{T}{^R_R^R_R^R_R^R_R}\phantom{\bigg)}$ has $x=0$, $\phantom{\big(}\tensor{T}{^R_R^R_R^R_R^R_L}\phantom{\big)}$ has $x=-2$, and $\tensor{T}{^R_L^R_L^R_L^R_L}$ has $x=-8$.

    For $C_3$ rotational symmetry, the response tensor $T$ is unchanged under rotations by $\Theta=2\pi/3$. For this to hold, it can only have non-zero tensor elements with $x\in\left\{0,\pm6\right\}$. On the other hand, the elements with $\nu=\nu'=R$ are $\tensor{T}{^R_R^R_R^m_n^o_p}$, for which $x\in\{0,\pm2,\pm4\}$, and similar for $\nu=\nu'=L$. Therefore, the only non-zero elements contributing to the response to circularly polarized input in a system with $C_3$ rotational symmetry are those elements with $x=0$, which are unchanged for any rotation angle $\Theta$, thus proving that such a coincidence response must have continuous rotational symmetry.

\section{Rotated Coincidence From Energy Levels}
\label{app:rotatedDetails}

    Looking at the energy levels in the one and two photon sectors, particularly focusing on the ground states $\ket{gs,1}$ and $\ket{gs,2}$, is a great way to understand features of the single-polarization and the unrotated two-polarization coincidence responses. Unfortunately, the energy levels alone do not explain the dramatic differences between the unrotated and rotated two-polarization coincidence responses. In order to understand those, we would need to include the filter applied to outgoing photons. 

    Consider first how incoming right-circularly polarized photons $\nu=R$ will populate the energy levels in Fig. \ref{fig:model_and_phase}(c). To leading order, the first photon drives the cavity from $\ket{gs,0}$ to $\ket{gs,R}$, and the second photon drives from $\ket{gs,R}$ to $\ket{gs,RR}$. To leading order, a filter for $\mu=R$ will transmit the full signal from these states in the same way as if there was no polarization involved. On the other hand, a filter for left-circularly polarized photons $\mu=L$ will entirely block any signal emitted by these states. If we want to understand the rotated transmission, we need to perform a detailed analysis of the states $\ket{gs,R}$ and $\ket{gs,RR}$, which contain fluctuations from $\DMat{LR}$ mixing the $R$ and $L$ character of each state. A perturbative expansion gives
    \begin{align}
        \ket{gs,R}=\ket{gs}\otimes\ket{R}&-\sum_{i\neq gs}\frac{\bra{i}\DMat{RR}\ket{gs}}{E_i-E_0}\ket{i}\otimes\ket{R} \\
        &-\sum_{i\neq gs}\frac{\bra{i}\DMat{LR}\ket{0}}{E_i-E_0}\ket{i}\otimes\ket{L}+\mathcal{O}(D^2). \notag
    \end{align}
    Rotated one-photon transmission will depend on the $\ket{i}\otimes\ket{L}$ terms, from which we get a factor of $\DMat{LR}$. A similar expansion for the two photon states finds that $\ket{i}\otimes\ket{LL}$ enters with two factors of $\DMat{LR}$. Thus, after filtering for rotated photons, the contribution from $n$-photon ground states no longer dominates by an order of magnitude for $\wIC\gtrsim0$ and instead we always get two powers of $D$ per photon as in Eq. (\ref{eq:rotatedCoincidencePerturbative}).

\end{document}